\documentclass{esub2acm}
\begin{document}

\setlength{\oddsidemargin}{1em}
\setlength{\evensidemargin}{1em}
\setlength{\textwidth}{160mm}
\setlength{\textheight}{220mm}

\include{times-roman.sty}


 \newtheorem{Theorem}{Theorem}
 \newtheorem{Property}{Property}
 \newtheorem{Proof}{Proof}
 \newtheorem{Lemma}{Lemma}
 \newtheorem{Proposition}{Proposition}
 \newtheorem{SubProof}{Proof}[Proof]
 \newtheorem{Definition}{Definition}
 \newtheorem{Axiom}{Axiom}
 \newtheorem{Condition}{Condition}
 
\input{psfig.tex}

\newcommand{\lmset}{\{\!|}
\newcommand{\rmset}{|\!\}}
\def \mset#1{{\lmset \! #1 \! \rmset}}
\def\spades{\spadesuit}
\def \trace#1{\langle #1\rangle}  
\let \ge        \geqslant
\let \le        \leqslant
\newcommand{\probthen}{\mapsto}
\newcommand{\until}{{~~\cal U~}}
\newcommand{\Prob}{\mathit Pr}
\newcommand{\prob}{\mathit prob}
\def \defeq {\stackrel{\mathrm{def}}{=}}
\def \undef {\perp}
\def\always{\ensuremath{\Box}}
\def\blackright{\mbox{$\;\;\-\!\!\!-\!\!\!\!-\!\!\!\triangleright$}}
\def \nat   {\mathbf{N}}
\def \real   {\mathbf{R}}
\def \power  {\mathbf{P}}
\def \union  {\bigcup}
\def \inter  {\bigcap}
\def \hide   {\backslash}

\def\mapsleft {\leftarrow\!\!\dashv}

\newcommand{\bc}{\begin{com}}
\newcommand{\ec}{\end{com}}

\newenvironment{com}[1]{\textmd{**\textsf{#1}**}}

\newsavebox{\savepar}
\newenvironment{boxit}{\begin{lrbox}{\savepar}
  \begin{minipage}[b]{5.2in}}
  {\end{minipage}\end{lrbox}\fbox{\usebox{\savepar}}}

\def\drop{\array[t]{@{}l@{}}}
\let\enddrop=\endarray
\def\ddrop{\array[t]{@{}ll@{}}}
\let\endddrop=\endarray

\title{Stochastic Model Checking for Multimedia}


\author{Jeremy Bryans, Howard Bowman and John Derrick\\
Computing Laboratory, University of Kent, Canterbury, CT2 7NF,
 UK.\\ (Phone: + 44 1227 764000, Email: J.Derrick@ukc.ac.uk.)}

\begin{abstract}
Modern distributed systems include a class of applications in which
non-functional requirements are important. In particular, these
applications include multimedia facilities where real time constraints
are crucial to their correct functioning. In order to specify such
systems it is necessary to describe that events occur at times given
by probability distributions and  stochastic automata have emerged as a
useful technique by which such systems can be specified and verified.

However, stochastic descriptions are very general, in particular they
allow the use of general probability distribution functions, and
therefore their verification can be complex.  In the last few years,
model checking has 
emerged as a useful verification tool for large systems.    In this paper we
describe two model checking algorithms for stochastic automata.  These
algorithms consider how properties written in a simple probabilistic
real-time logic can be checked against a given stochastic automaton.
\end{abstract}
\begin{bottomstuff}
  \begin{authinfo}
    \name{J. Bryans} 	  
   \address{Computing Laboratory, University of Kent at Canterbury,
    Canterbury, Kent, CT2 7NF  PHONE: +44 1227 827697} 	  
    \name{H. Bowman} 	  
    \address{Computing Laboratory, University of Kent at Canterbury,
    Canterbury, Kent, CT2 7NF  PHONE: +44 1227 823815} 	  
    \name{J. Derrick} 	  
    \address{Computing Laboratory, University of Kent at Canterbury,
    Canterbury, Kent, CT2 7NF  PHONE: +44 1227 827570} 	  
  \end{authinfo}
\permission
\end{bottomstuff}

\markboth{J. Bryans, H. Bowman and J. Derrick}{Stochastic
Model-Checking for Multimedia}
\maketitle

\section{Introduction}

In this paper we describe and compare two model checking algorithms
for stochastic automata.  The reason for building such model checking
algorithms is to support the verification of non-functional properties
in distributed multimedia systems.

The advent of distributed multimedia applications such as video
conferencing, collaborative virtual environments, video on demand etc,
place great demands on the specification and design of such systems
because of the need to describe and verify non-functional
requirements~\cite{tempo-book}. These non-functional requirements
typically involve real time constraints such as placing bounds on
end-to-end latency, and are often called {\it Quality of Service
(QoS)}~\cite{tempo-book} requirements because they reflect the overall
quality of delivery as opposed to the functional aspects.

In order to specify and verify such constraints it is necessary not
only to be able to describe deterministic timing concerns but also
probabilistic and stochastic systems. That is, in practice timings
cannot be assumed to be fixed (deterministic timings) but events can
occur at different times with particular probabilities. Therefore it
is necessary to describe timings that occur according to certain {\it
probability distributions}.  For example, in a network specification
it is not sufficient to assume that the packet deliveries arrive at
fixed predetermined times, instead we need to model the system where
they might arrive at times determined by (for example) an exponential
distribution. 

There are now a number of techniques which can be used to describe
such systems, e.g. Queueing Systems~\cite{Kleinrock-1-1975},
Generalised Stochastic Petri-nets
\cite{Ajmone_Marsam-Conte-Balbo-1984}, Markov
Chains~\cite{Stewart-1994}, generalised semi-Markov
processes~\cite{Glynn-1989}, Stochastic Process
Algebra~\cite{Hillston-1996} and Stochastic
Automata~\cite{D'Argenio-1999} etc. In this paper we consider
Stochastic Automata (which are related to timed
automata~\cite{Alur-Dill-1994}). We define two {\em model
checking}~\cite{Baier-Kwiatkowska-1998} algorithms for them.

Stochastic automata are a very promising specification and
verification paradigm.  In particular they allow the study of both
functional and non-functional requirements within the same
description, giving a more complete view of overall performance than,
say, a queueing theory description of the problem.  They also support
not just exponential distributions but general distributions. The
issue here is the following. In a stochastic specification we need to
associate a distribution function $F$ with an action $a$ so that we
can describe the probability of the time delay after which $a$ can
happen.  Stochastic automata naturally allow general distributions, in
contrast say to stochastic process algebras which usually restrict
themselves to exponential distributions~\cite{Hillston-1996}.

 In practice it is unrealistic to only consider exponential
distributions and it is necessary  for arbitrary  distributions
(e.g. uniform, gamma, deterministic etc) to be considered. 
For example, it is often assumed that packet lengths are exponentially
distributed. However, in reality this is not the case,
rather they are either of constant length (as in ATM cells~\cite{Tanenbaum96})
or they are uniformly distributed with minimum and maximum size (as in Ethernet
frames~\cite{Tanenbaum96}). Stochastic automata allow such arbitrary
distributions 
to be used.

There are ostensibly two ways to move from the tractable case of
exponential distributions to the less tractable case of generalised
distributions.  One approach is to make small generalisations of
markov chains by allowing limited forms of non-memoryless behaviour
(see e.g. GSPNs~\cite{Ajmone_Marsam-Conte-Balbo-1984}).  However, the
problem with this approach is that there will 
always be classes of distributions that cannot be modelled.  The
alternative is to allow any distribution, but then use heuristics and
coarse approximation techniques to contain the problem of
intractability.  The majority of work on this topic follows the first
of these approaches.  However here we investigate the feasibility of
the second approach and thus we impose few constraints on the
generality of the distributions we allow in our stochastic automata.  

\begin{figure*}[t]
\begin{center}
   \ \psfig{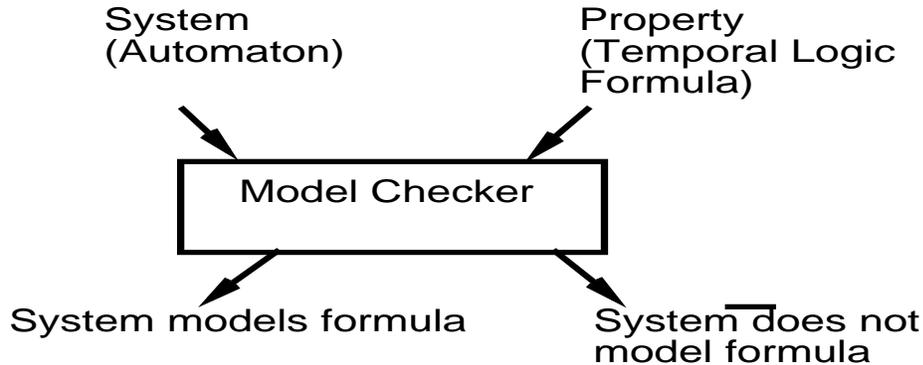}
 \end{center}
 \caption{Model checker}
\label{fig:model-checker}
 \end{figure*}

Because stochastic automata are related to  timed automata, verification
strategies for stochastic automata can be derived by using the extensive work
on verification for timed automata, see e.g.~\cite{Uppaal}
\cite{Kronos} \cite{Hytech}. 
The particular verification technique we consider is model
checking~\cite{Alur-Courcoubetis-Dill-1990}.  This is perhaps the most
successful 
technique to have arisen from concurrency theory.  The basic approach
is to show that an automaton description of a system satisfies a
temporal logic property, see  Figure~\ref{fig:model-checker}.

In accordance with a number of other workers,
e.g.~\cite{Baier-Katoen-Hermanns-1999}, we view the application of
model checking to analysis of stochastic systems as a very exciting
combination, since it provides a form of generalised transient
analysis --- for example the property $[ \lnot error \!
\until_{\!\!\!<1000} \; error] < 0.01$ states that the probability of
first reaching 
an {\em error} state within 1000 time units is less than 1 percent,
and whether a particular stochastic system satisfies this property can
be investigated.

In defining our model checking algorithm we draw heavily on the
experience of model checking timed automata
e.g.~\cite{Uppaal}. However, the move from timed to stochastic leads
to new issues that must be tackled. In particular, many of the
properties that we wish to verify are inherently probabilistic.
Conventional model checking allows us to answer questions such as ``Is
a particular sequence of events possible?'', but in stochastic model
checking we want to ask ``What is the probability of this sequence of
events?''.  To do this we will check an automaton against a simple
probabilistic temporal 
logic.

We present two approaches to model checking stochastic automata.  Both
approaches are enumerative in the sense that, in showing whether a
property holds, they enumerate reachable configurations of the
system.  However, the methods by which they determine the probability
of being in a particular configuration are quite different.
Specifically, one derives probabilities by integrating the relevant
probability density functions, while the second responds to the
difficulties incurred in evaluating these integrals (which will become
clear during the paper) by employing a discretisation process.

The structure of the paper is as follows. In
Section~\ref{sec:stochastic-automata} we introduce stochastic automata
illustrated by a simple example.  In Section \ref{sec:logic} we define
a small probabilistic real-time logic, in which we can express simple
properties that we wish to check our stochastic automata against.  The
first algorithm is presented in Section~\ref{sec:alg-one}, and the
second is presented in~\ref{sec:alg-two}.  Section~\ref{sec:example}
looks at an example of the operation of the second algorithm and
Section~\ref{sec:proof} considers some issues of correctness and
convergence relating to the second algorithm, and
Section~\ref{sec:complexity} looks at the time and space complexity.
We conclude in 
Section~\ref{sec:conclusions}.

%
%
%

\section{Stochastic Automata} \label{sec:stochastic-automata}

In this section we introduce stochastic automata using a small
 example.  Stochastic automata are related to timed
 automata~\cite{Alur-Dill-1994}, however stochastic clock settings are
 used, instead of the strictly deterministic timings used in timed
 automata.  We begin with the formal definition of stochastic
 automata, then present a simple example.  We use the definition of
 stochastic automata presented
 in~\cite{D'Argenio-Katoen-Brinksma-1998}.
\begin{Definition}{\upshape
A {\em stochastic automaton} is a structure $({\cal S},s_0,{\cal
  C},{\bf A},\blackright,\kappa,F)$ where: 
\begin{itemize}
\item[$\bullet$] ${\cal S}$ is a set of {\em locations} with $s_0 \in
{\cal S}$ being the 
  {\em initial location}, ${\cal C}$ is the set of all {\em clocks},
  and {\bf A} is a set of {\em actions}.
\item[$\bullet$] $\blackright \subseteq {\cal S} \times ( {\bf A} \times
  {\cal P}_{\mathrm{fin}}({\cal C})) \times {\cal S}$ is the set of
  {\em edges}.  If $s$ and $s'$ are states, $a$ is an action and $C$
  is a subset of ${\cal C}$, then we 
  denote the edge $(s,a,C,s') \in \blackright$ by $s
  \stackrel{a,C}{\blackright} s'$ and we say that $C$ is the {\em
  trigger set} of action $a$.  We use   $s \stackrel{a}{\blackright}
  s'$ as a 
  shorthand notation  for $\exists C.s \stackrel{a, C}{\blackright} s'$.

\item[$\bullet$] $\kappa : {\cal S} \rightarrow {\cal P}_{\mathrm
{fin}}(C)$ is the {\em clock setting function}, and indicates which
clocks are to be set in which states, where ${\cal
P}_{\mathrm{fin}}({\cal C})$ is the finite powerset of clocks.

\item[$\bullet$] $F: {\cal C} \rightarrow ({\cal R} \rightarrow [0,1])$
assigns to each clock a {\em distribution function} such that, for any
clock $x$, $F(x)(t) = 0$ for $t < 0$; we write $F_x$ for $F(x)$ and
thus $F_x(t)$ states the probability that the value selected for the
clock $x$ is less than or equal to $t$.
\end{itemize}
Each clock $x \in {\cal C}$ has an associated random
variable with distribution $F_x$.  
To facilitate the model checking, we introduce a function $\xi$ which
associates locations with sets of atomic propositions.
\[
\xi:  {\cal S} \mapsto {\cal P}(AP)
\]
where AP is the set of atomic propositions.  $\hfill\Box$ }\end{Definition}

It is necessary to impose some limitations on the stochastic automata
which can be used with the model checking algorithms.  In particular,
we require that each clock distribution function $F_x$ must have a
positive finite upper bound and a non-negative lower bound, and must
be continuous between these bounds.  The finiteness constraints mean
that there are certain distribution functions which we must
approximate.  We further assume that clocks are only
used on transitions emanating from  states in which they are set.


 \begin{figure*}
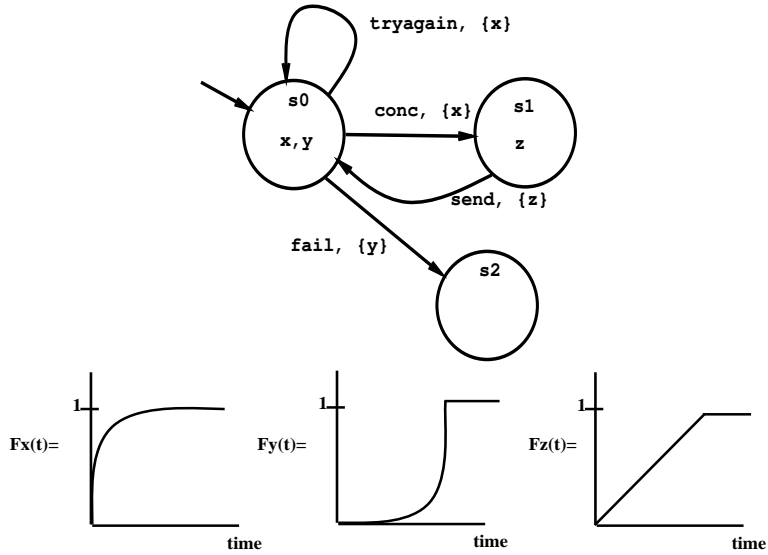

 \begin{center}
   \ \psfig{figure=stoch-automaton2.pic,height=2in,width=2in}
 \\
   \ \psfig{figure=three-graphs.pic,height=1in,width=4in}
 \end{center}
 
 \caption{The packet producer} \label{fig:stoch-automaton}
 \end{figure*}
 
As an example, consider the simple packet producer (which is a
component in  a large multimedia specification) 
in Figure~\ref{fig:stoch-automaton}.  This is written 
\[
\begin{array}{l}
(\{s_0,s_1,s_2\},s_0,\{x,y,z\}, \\ 
\{tryagain, conc, send, fail\}, \blackright,\kappa,\{F_x,F_y,F_z\})
\end{array}
\]
where 

\[
\begin{array}{rcl}
\blackright & = & \{(s_0,tryagain,\{x\},s_0), (s_0,conc,\{x\},s_1), \\
& & (s_1,send,\{z\},s_0), (s_0,fail,\{y\},s_2)\} \\
\end{array}
\]
\[
\begin{array}{ccc}
\kappa(s_0)  =  \{x,y\}, &
\kappa(s_1)  =  \{z\}, &
\kappa(s_2)  =  \{\} 
\end{array}
\]
and the 
distribution functions for clocks $x$, $y$ and $z$ are 
\[
\begin{array}{rcl}
F_x(t)  & = & 2t-t^2, \mathrm{if~~} t\in[0,1] \\
        & = &      0, \mathrm{if~~}  t < 0 \\
        & = &    1, \mathrm{otherwise} \\
\end{array}
\]
\[
\begin{array}{rcl}
F_y(t) & = & t^2, \mathrm{if~~}  t\in[0,1] \\
       & = &       0,  \mathrm{if~~}  t < 0  \\
       & = &      1,      \mathrm{otherwise}
\end{array}
\]
and
\[
\begin{array}{rcl}
F_z(t) & = & t, \mathrm{if~~}  t\in[0,1] \\
       & = &       0,  \mathrm{if~~}  t < 0  \\
       & = &      1,      \mathrm{otherwise}
\end{array}
\]
 as depicted. The horizontal axis measures time, and the vertical
axis measures the probability of the clock being set to a value less
than that time. 

The packet producer starts in location $s_0$, and attempts to
establish a connection with its medium.  Three options are possible at
this stage.  Either the medium allows a connection, the medium tells
the packet producer to try again or the medium takes too long and the
connection fails (is timed out).  These options are modelled in the
automaton by setting clocks $x$ and $y$ according to the functions
$F_x$ and $F_y$.  If clock $x$ expires first then there is a
nondeterministic choice between the transition labelled  {\em conc}
(which moves the automaton to state $s_1$)  and the transition labelled 
{\em tryagain} (which moves the automaton back to state $s_0$).  This
choice is nondeterministic because in reality it would depend on 
the medium, which we have not specified here.  If clock $y$ expires
first, then action {\em fail} is triggered (we say that $\{y\}$ is the
{\em trigger set} of {\em fail}) and the automaton moves to state
$s_2$.  This corresponds to the medium taking too long to respond, and
nothing further happens.     

This example has been chosen because it is small enough that we can
show, in their entirety, the set of configurations that our model
checking algorithms enumerate.  Thus it can be used to illustrate our
two algorithms.  But, in addition, we have chosen it because it is
canonical in the sense that it illustrates the key concepts of
stochastic automata, e.g. simultaneous enabling of multiple
transitions generating non-determinism.  The reader should also notice
that this is a good example of a situation in which steady-state analysis
is not interesting.  Specifically, in the steady state, all the
probability mass will be in state $s_2$.  Thus, the sort of questions
we wish to ask about such  a system are about its transient behaviour,
e.g. what is the probability of reaching state $s_2$ within a
particular period of time  and indeed this is exactly the type of
question we will be able to formulate with the logic we introduce in
the next section and answer with our model checking algorithms. 

\section{A Probabilistic Real-Time Temporal Logic} 
\label{sec:PRTL}
\label{sec:logic} 

\subsection{The Logic}

In this section, we introduce a simple probabilistic temporal logic.
The purpose of the logic is to express properties that we wish to
check the stochastic automaton against.  The logic we define allows us
to check a range of such properties.

Recall that the region tree contains nondeterminism, and so we resolve
this using the notion of {\em adversaries} (see for
example~\cite{Baier-Kwiatkowska-1998}).  An adversary of a stochastic
automaton can be thought of as a scheduler, which resolves any
nondeterministic choices which the stochastic automaton must make.  An
adversary may vary it's behaviour according to the previous behaviour
of the automaton, or it may prescribe that for all non-deterministic
choices a particular branch is always preferred.
See~\cite{D'Argenio-1999} for examples of adversaries.

We assume that when we wish to model check a property against an
automaton, we are also given an adversary to resolve the
nondeterminism within the automaton. (Without this adversary,
enumerative analysis would not be possible; the provision of an
adversary is a prerequisite of model checking.)  We can now, for
example, answer such questions as ``Given a stochastic automaton and
an adversary, is the probability of a {\em send} event occurring
within 5 time units greater than 0.8?''.

The syntax of our logic is

\[
\psi ::=   \mathsf{tt} \mid \mathsf{ap} \mid \lnot \psi \mid \psi_1 \land
\psi_2 \mid [\phi_1  \until_{\!\!\sim c}\, \phi_2] \simeq p
\]
\[
\phi ::= \mathsf{tt} \mid \mathsf{ap} \mid \lnot \phi \mid \phi_1 \land \phi_2
\]

where $[\phi_1 \until_{\!\!\sim c}\, \phi_2] \simeq p$ is a {\em path
formula}.  The path formulae can only be used at the top level ---
they cannot be nested.  This is because the model checking algorithm
we give can only evaluate path formulae from the initial state and is
a necessary restriction of the current approach.
Further: $c \in {\nat}$ (natural numbers), {\sf ap} is an atomic
proposition, {\sf p} $\in [0,1]$ is a probability value and $\simeq, \sim
\in\{<,>,\leq,\geq\}$. 

We can define a number of derived operators.  For example, other
propositional operators are defined in the usual way:-

\[
\begin{array}{rcl}
\mathsf{ff} & \equiv & \lnot \mathsf{tt} \\
\phi_1 \lor \phi_2 & \equiv & \lnot (\phi_1 \land \phi_2) \\
\phi_1 \Rightarrow \phi_2 & \equiv & \lnot \phi_1 \lor \phi_2
\end{array}
\]
and we can define a number of abbreviations of a number of temporal
operators. 
\[
\begin{array}{rcl}
[\Diamond_{\sim c}\phi] \simeq p & \equiv & [\mathsf{tt} \!\!\until_{\!\!\sim
c}\;\phi] \simeq p \\ 
\hspace{1pt}[\always_{\sim c} \phi] \simeq p & \equiv & [\lnot
\Diamond_{\sim c}\lnot \phi] \simeq p  \\
\hspace{1pt}[\always \phi] \simeq p & \equiv & [\always_{\geq 0} \phi]
\simeq p \\%
\hspace{1pt} [\Diamond \phi] \simeq p & \equiv & [\Diamond_{\sim 0}
\phi] \simeq p \\
\hspace{1pt} \forall [\phi_1 \!\!\until_{\!\!\sim c} \phi_2] & \equiv &
[\phi_1 \!\!\until_{\!\!\sim c} \phi_2] = 1 \\
\hspace{1pt} \exists [\phi_1 \!\!\until_{\!\!\sim c} \phi_2] & \equiv &
[\phi_1 \!\!\until_{\!\!\sim c} \phi_2] > 0 \\
\hspace{1pt}\forall\always \phi  & \equiv & \forall[\always \phi] \\%
\hspace{1pt}\exists\always \phi  & \equiv & \exists[\always \phi] \\%
\hspace{1pt}\forall\Diamond \phi  & \equiv & \forall[\Diamond \phi] \\%
\hspace{1pt}\exists\Diamond \phi  & \equiv & \exists[\Diamond \phi] \\%
\end{array}
\]

where $\forall$ and $\exists$ are the branching time temporal logic
operators, {\em for all} and {\em exist}~\cite{Emerson-1990}.
See~\cite{Baier-Katoen-Hermanns-1999} for similar definitions. 

With this syntax, an example of a valid formula that we can check
would be $[\mathsf{tt}\!\!\until_{\!\!< 10}\, \mathsf{send}] > 0.8$ which
says that the probability of reaching a {\sf send} event within 10
time units is greater than 0.8.

\subsection{Model Checking}

It should be clear that since we do not allow temporal formulae to be
nested we can use the following recipe in order to model check a
formula $\psi$ of our logic against a stochastic automaton $A$.

\begin{itemize}
\item[1.] For each until subformula (i.e. of the form $[\phi_1 \!\!
\until_{\sim c} \phi_2] \simeq p$) in $\psi$ perform an individual
model check to ascertain whether 
\[
A \models [\phi_1 \!\!\until_{\!\!\sim c}\phi_2]\simeq p
\]
\item[2.] Replace each until formula in $\psi$ by {\sf tt} if its
corresponding model check was successful, or {\sf ff} otherwise.
\item[3.] Replace each atomic proposition in $\psi$ by {\sf tt} or {\sf
ff} depending upon its value in the initial location of $A$.  
\item[4.] $\psi$ is a now ground term, i.e. truth values combined by a
propositional connective ($\lnot$ and $\land$).  Thus, it can simply
be evaluated to yield a truth value.  The automaton is a model of
$\psi$ if this evaluation yields {\sf tt}, and is not otherwise.  
\end{itemize}

This recipe employs standard techniques apart from the individual
checking that $A \models [\phi_1 \!\!\until_{\!\!\sim c}\phi_2]\simeq
p$ and this is what our two algorithms address.

\section{The Region-tree Algorithm}
\label{sec:alg-one}

In this section we introduce the first algorithm.

In model checking, we take a temporal logic predicate and seek to
establish whether it is true for our particular specification.  For
example, we might try to establish whether the above stochastic
automaton has the following property: Is the probability that a packet
will be successfully sent within ten time units greater than
eighty percent? In order to do this, we need to  define a means by
which we can check the stochastic 
automaton against this logic.  To achieve this the temporal logic and
the specification must have the same semantic model.
In~\cite{D'Argenio-Katoen-Brinksma-1998}, stochastic automata are
given a semantics in terms of {\em probabilistic transition systems},
and so the temporal logic is given a semantics in terms of
probabilistic transition systems as well, see
Appendix~\ref{sec:stoch-semantics}.

\subsection{Region Trees} \label{sec:region_trees}

For practical purposes, however, we cannot construct the probabilistic
transition system, since it is an infinite structure, (both in
branching and depth.)  We instead
construct a {\em region tree} from the specification.  This is
finitely branching, but may be infinite in depth.  Thus, a particular
region tree represents an unfolding of the stochastic automaton to a
certain depth.  In fact, we use the
temporal logic formula to construct a probabilistic region tree, which
is  used to verify the temporal logic formula.  More precisely, the
region tree is expanded until sufficient probability has been
accumulated to ascertain the truth or falsity of the formula (this
will become more clear shortly.)  In this section,
we describe how to construct  region trees from
stochastic automata.


 We begin with the definition of a valuation, which we use to record
 the values of all the clocks in a particular state at a particular
 moment in time.  The unique clock $a \in {\cal C}$, which we add to
 the set of clocks, is used  to facilitate the model checking.  It
 keeps track of the total time elapsed in the execution of the
 stochastic automaton, but plays no part in the behaviour of the automaton.

\begin{Definition}
  A {\em valuation} is a function $v:{\cal C}\bigcup\{a\} \rightarrow
  {\cal R} \bigcup \{\perp\}$ such that $v(x) = \bot$ or $v(x)
  \leq x_{max}$, where 
  $x_{max}$ is the maximum 
  value to which clock $x$ can be set.    If $d \in {\cal R}_{\geq 0}$,
  $v-d$ is defined by $\forall 
  x \in {\cal C}\bigcup\{a\}. (v-d)(x) \defeq v(x)-d$.  
The function $\min(v)$ returns the value of the smallest defined
clock.    $\hfill\Box$
\end{Definition}
 
Since we assume that clocks are only used in the states in which they
are set, there is no need to remember their value once the state has
been exited.  Only the clock $a$ maintains its value; the rest are set
to $\bot$.  At the initialisation of a stochastic automaton, clock $a$
is set to some natural number, (we will show later how we choose this;
it depends on the formula we are interested in) and all other clocks
are undefined.  We define this initial valuation as ${\mathbf O}_n$,
if ${\mathbf O}(a) = n$.

We also need a notion of equivalence between the valuations, which
will enable us to construct the  regions within the
probabilistic region tree.  The issue here is the following.  Although
the size of the tree will be potentially infinite, at each node we
wish to have {\em finite} branching.  We achieve this because, although
there are an infinite number of valuations possible for any particular
state, there are a finite number of valuation equivalence classes.
This gives us the finite branching.  

\begin{Definition}
Two clock valuations  $v$ and $v'$ are {\em equivalent}  (denoted $v
\cong v'$) provided the following conditions hold:

\begin{itemize}
\item For each clock  $x \in {\cal C}\bigcup\{a\}$, either both $v(x)$
  and $v'(x)$ 
  are   defined, or $v(x) = \undef$ and $v'(x) = \undef$.   
\item For every (defined) pair of clocks $x,y \in {\cal
    C}\bigcup\{a\}.v(x) < v(y) \iff 
  v'(x) < v'(y)$.   
\end{itemize}  

The same clocks are defined in each valuation, and the order of the
values of the defined clocks is all that is important.    $\hfill\Box$
\end{Definition}

The reason that the order of the values of the defined clocks is all
that is important in the definition of a valuation equivalence class
is that the actions are triggered by the first clock to expire.
Therefore we only need to know whether one clock is greater than or
less than another.   Also note that there is a probability of zero
that different clocks are set to the same value.  This is because all
distributions are assumed to be continuous.

\begin{figure*}[t]
\begin{center}
  \ \psfig{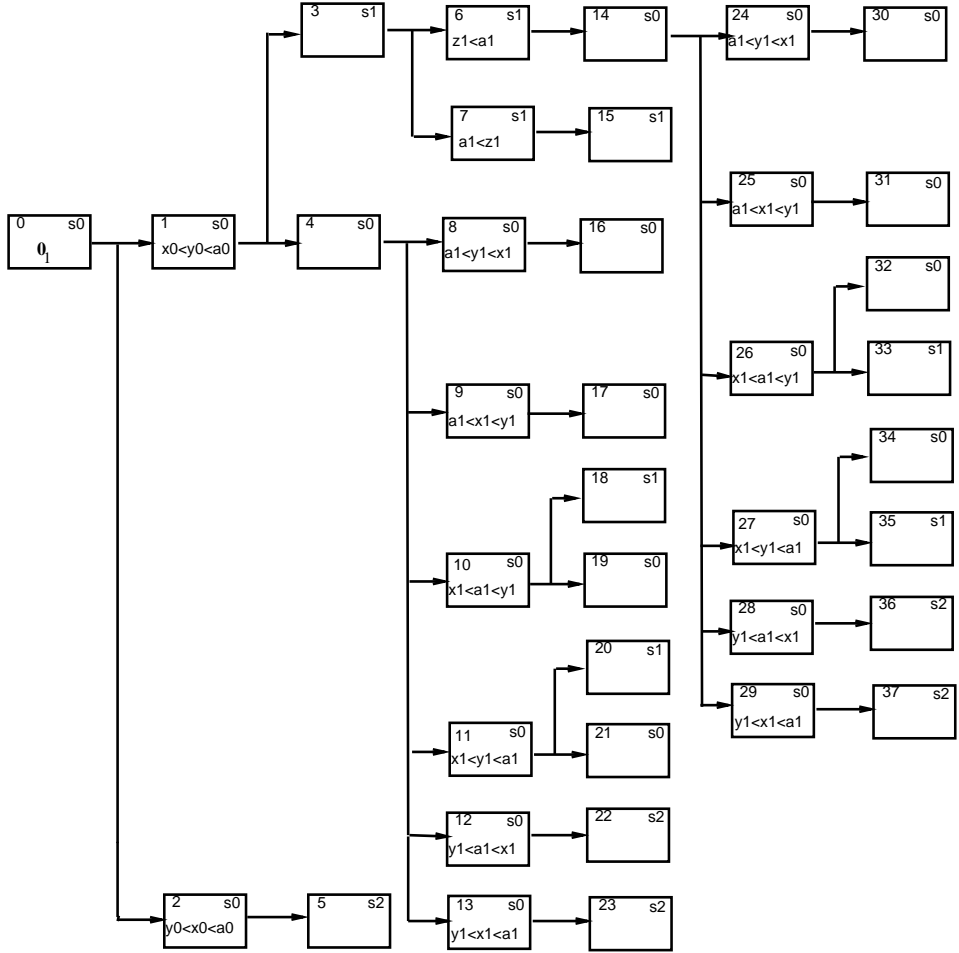}
\end{center}
\caption{The region tree}
\label{fig:region-tree}
\end{figure*}

We are now in a position to describe how a region tree is constructed
from a stochastic automaton. 
Intuitively, we build the region tree by ``unfolding'' the stochastic
automaton.  At each newly reached state, we calculate all possible
valuations (up to $\simeq$) and the probabilities of each one, then
from each of these 
(state,valuation) pairs we calculate the possible new states and
repeat.

Suppose we wish to construct the region tree for the
stochastic automaton in Figure~\ref{fig:stoch-automaton}.

The resulting region tree (up to a particular level of unfolding) is
given in Figure~\ref{fig:region-tree}.  The first node is labelled
with the location $s_0$, where the SA starts, the valuation ${\mathsf
0}_1$, (i.e.  $(1,\perp,\perp)$) since clocks $x$ and $y$ have not yet
been set, and clock $a$ is set to value one.  Clock $a$ is set
according to the time value on the formula in which we are interested;
we will give the example formula in Section~\ref{sec:prt}.  The clocks
$x$ and $y$ are then set, giving a potential $3! = 6$ different
equivalence classes.  However, these can be reduced to two by
observing that clock $a$ will be fixed on 1 and $x_{max} = y_{max} =
1$ and the probability of either $x$ or $y$ being set to exactly 1 is
zero\footnote{This coincidence of $a$, $x_{max}$ and $y_{max}$ is
assumed in order to simplify our presentation; the next iteration
illustrates the general case.}.  Using the convention that we
subscript the clock variables by the iteration number, in order to
distinguish different settings of the same clock, the two possible
equivalence classes are therefore $v_0(y) < v_0(x) < v_0(a)$ and
$v_0(x) < v_0(y) < v_0(a)$, where $v_0(a) = 1$ in both cases.  For
convenience, we will write $x_0$ for $v_0(x)$, $y_0$ for $v_0(y)$
and $a_0$ for $v_0(a)$.

If clock $x$ is set to less than clock $y$, the automaton will allow
time to pass in location $s_0$, and each clock will count down, until
clock $x$ reaches zero.  Then, either action {\em tryagain} or action
{\em conc} will fire (the choice is nondeterministic), and the
automaton will enter location $s_0$ or $s_1$ respectively.  The time
at which this occurs will obviously vary according to the initial
value of the clock $x$.  The possible locations entered are depicted
by regions 3 and 4 in the region tree in Figure~\ref{fig:region-tree},
where clocks $x$ and $y$ (since they are irrelevant in these regions)
are not recorded.  The initial value of clock $a$ when moving from
region 1 to either region 3 or region 4 will be $1 - x_0$ (we will
denote this value as $a_1$).  Thus,
it will be in the range $(0,1)$.

If clock $y$ is set to less than clock $x$ (represented by region 2),
 then the action {\em fail} fires, 
causing the automaton to enter location $s_2$, and this is depicted
by region 5 in the region tree.  Again, all we can say about the value
of clock $a$ at this stage is that it lies in the range $(0,1)$.

From region 3 there are two possibilities.  Either clock $z$ is set to
less than $a_1$, (region 6), or it is set to greater than $a_1$
(region 7).  From region 6 the action {\em send} will occur before the
clock $a$ expires, moving the automaton to location $s_0$ and the
region tree to region 14. From region 7 the clock $a$ will expire
before the action {\em send} occurs.  The region tree moves to region
15, and the automaton remains in state $s_1$.

From region 4 (location $s_0$) both clocks $x$ and $y$ are reset according to 
their probability density functions, to values $x_1$ and
$y_1$. Since we cannot now be 
sure about the value of clock $a$, we have $3!=6$
equivalence classes, and these are represented by regions 8 to 13 when
we unfold the SA another 
level.

In regions 8 and 9 $a_1$ is less than the (new) initial values of
clocks $x$ and $y$: these regions represent the case where clock $a$
expires before either of $x_1$ and $y_1$.  When we consider a
particular temporal logic formula this will represent the case where
time has run out, and so the region tree moves to either region 16 (if
$y$ expired 
first) or  region 17 (if clock $x$ expired first).

Regions 10 and 11 represent the valuation equivalence classes where
$x_1$ is less than both $a_1$ and $y_1$, and so from these
clock $x$ will expire first, either action {\em tryagain} or {\em
conc} will be performed, and the stochastic automaton will enter
either location $s_0$ or $s_1$ (regions 18---21).

 Region 12 and represent the valuation equivalence classes where
$y_1$ is less than both $a_1$ and  $x_1$, so clock $y$ will expire
first, action {\em fail} 
will fire, and the automaton will enter location $s_2$ (region
22---23).

The region tree can be expanded further if necessary.  There is no
need to continue to expand regions 5, 15, 16, 17, 22 and 23, because in
all of these either the clock $a$ has expired or the stochastic
automata has reached location $s_2$, which is a deadlocked state, and
there is no further 
information to be gained.  In Figure~\ref{fig:region-tree}, further 
regions are derived from region 14 in the same way as above; these are
needed when we build the probabilistic region tree in the next
section.  

\subsection{Probabilistic Region Trees} \label{sec:prt}

Given a stochastic automaton, adversary and formula $\psi =  [\phi_1\!\!
\until_{\!\!\sim c}\, \phi_2]\simeq {\tt p}$ the model
checking algorithm consists of a number of iterations which are
repeated until the formula is found to be either true or false. 

An iteration unfolds the region tree by expanding each leaf node.  At
each iteration stage there are two steps.  The first step resolves
the nondeterministic choices in the newly expanded region tree using
the given adversary.  The second step then calculates the
probabilities on each node in the newly expanded part of the tree.

%
%
%
%

\begin{figure*}[t]
\begin{center}
   \ \psfig{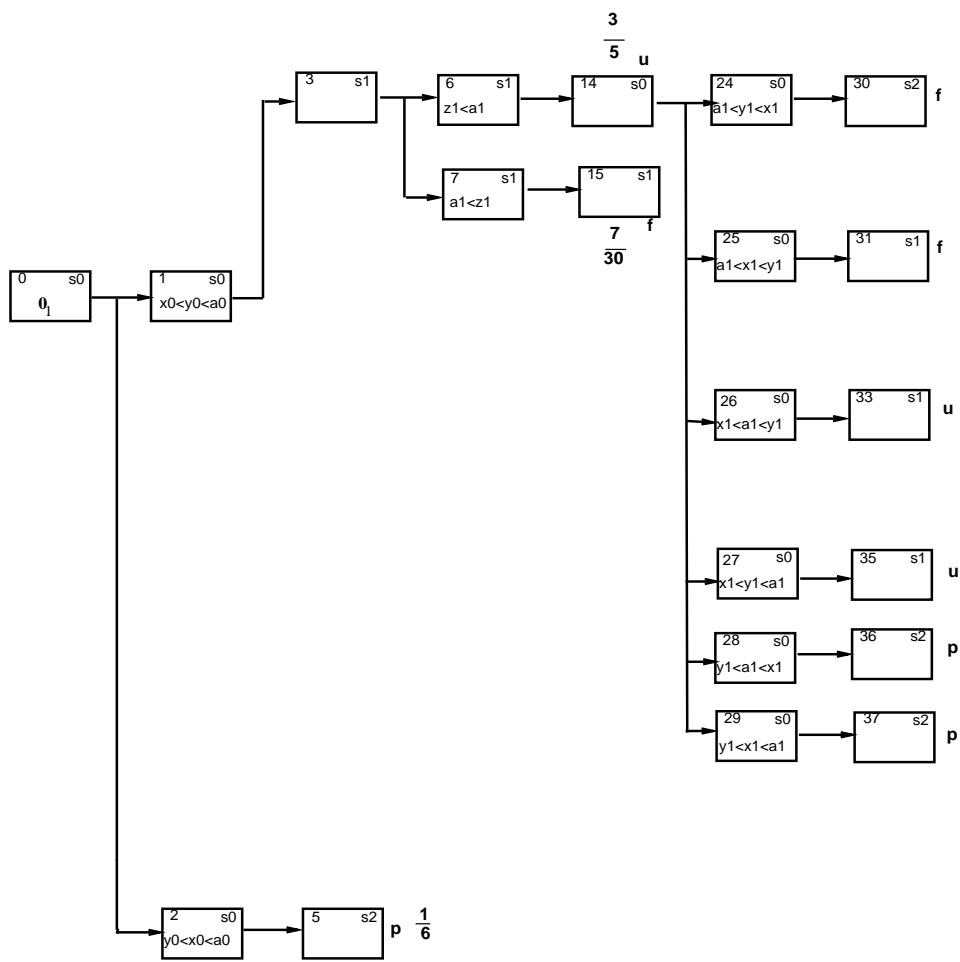}
 \end{center}
 \caption{The probabilistic region tree}
\label{fig:prob-region-tree}
 \end{figure*}

%
%
%
 
 The region tree (Figure~\ref{fig:region-tree}) represents an unfolding of the
 stochastic automaton without the nondeterministic choices being resolved.
 The probabilistic region tree (Figure~\ref{fig:prob-region-tree}) records the
 resolution of the nondeterministic choices and the probabilities at
 the final nodes represent the chances of taking the particular
 sequence of actions that end in that node.

At each iteration, we update the information we have on the
probability of a path satisfying the formula.  To do this, we define three
new propositions, and each node of the probabilistic region tree is
labelled with {\sf p}, {\sf f} or {\sf u}: {\sf p}, 
if it has {\it passed} (it is the end of a path which models the
bounded until formula $\psi$);
{\sf f}, if it has {\it failed} (it is the end of a path which cannot
model $\psi$), or {\sf u}, if it is {\it undecided}.  We also have two
global variables, $\Sigma {\sf p}$ and $\Sigma {\sf f}$, which keep
running totals of the probabilities of the {\it pass} and {\it fail}
paths.

The basic idea of the model checking algorithm is that we check the values of
$\Sigma {\sf p}$ and $\Sigma {\sf f}$ at each stage, and if we cannot
deduce from these the truth or falsity of the formula we are checking,
we look more closely at the undecided nodes.  That is,  we extend the
undecided paths by each possible 
subsequent 
action, label these new nodes {\sf p}, {\sf f} or {\sf u}, and
calculate their probabilities.  We then add these probabilities to
$\Sigma {\sf p}$ and $\Sigma {\sf f}$ and repeat. 

We will begin by demonstrating the technique for an example. The full
algorithm appears as appendix~\ref{sec:first-algorithm}.
Consider the example stochastic automaton
(Figure~\ref{fig:stoch-automaton}).

Let us consider the formula 
\[
\psi = [(\phi_0\lor \phi_1) \!\!\until_{\!\!<1}\; \phi_2] \geq 0.9
\]
 where $\phi_0$ (resp. $\phi_1$, $\phi_2$) is the proposition that we
are in state $s_0$ (resp. $s_1$, $s_2$).  The question\footnote{In fact, the
algorithm can easily be adapted to handle questions such as ``what is
the probability (to within some $\epsilon$) of a formula such as $[\phi_0\!\!
\until_{\!\!<1}\; \phi_2]$ being true?''.}  we are therefore asking is: is the
probability of reaching location $s_2$ (failing) within one time unit
greater than $0.9$?

Note that a steady state analysis will tell us only that the automaton
will fail (reach state $s_2$) eventually, but here we want to obtain
information about the transient behaviour of the automaton.
 The nondeterministic choice that has to
be made is between location $s_1$ and $s_2$.
We will consider  the benevolent adversary, i.e. the one that always
chooses location $s_1$.  
 
Consider region 1 first (Figure~\ref{fig:region-tree}).  It has two
possible outgoing transitions, 
and the choice between them is made nondeterministically.  So we must
refer to the adversary, which chooses location $s_1$, that is, region
3.   Region 4 is not generated.  We note that the value of clock $a$ 
is
greater than zero (so time has not run out), and 
that proposition $p_0 \lor p_1$ is true (so the temporal logic formula is
able to be satisfied), so this region is labelled with {\sf u} (undecided). 

In region 5 proposition $p_2$ is true, and clock $a$ is still greater
than zero, so this region is labelled as passed {\sf p}, and region 5
becomes a terminal node.

In region 6 $a_1$ is greater than the (new) initial value of
clock $z$, and therefore  the {\em send}
action will fire before the clock $a$ expires. The region is therefore
labelled {\sf u}.   

In region 7 $a_1$ is less than the (new) initial value of
clock $z$, and therefore time will run out before the {\em send}
action has a chance to fire. The region is therefore labelled {\sf f}.

From region 6 the {\em send} action moves the automaton to location
$s_0$ (region 14), and from here there are 6 possibilities for the
setting of the clocks.  

Regions 24 and 25 represent the valuation equivalence classes where
$a_1$ is less than $x_1$ and $y_1$.  Since clock $a$ will
expire before either clock $x$ or clock $y$,  we know that these paths
will not reach location $s_2$ in less than one time unit, so regions
30 and 31 will be labelled {\sf f}.  The remainder of the tree is
generated in a similar manner.  

Figure~\ref{fig:prob-region-tree} represents two unfoldings.  In
order to determine whether the formula is true we also have to
calculate the probabilities on the nodes.  If the sum of the {\em
  pass} and the sum of the {\em fail} nodes is sufficient to tell us
whether the formula is true then we can stop here, otherwise  we
unfold the tree another level.   



%
%

To determine the probabilities on the arcs, we need to use probability
density functions $\sf{P}\!_x$, $\sf{P}\!_y$ and $\sf{P}\!_z$ of the
functions $F_x$, $F_y$ and $F_z$, which we find by differentiating
$F_x$, $F_y$ and $F_z$ between
their upper and lower bounds and setting to zero everywhere else.

\[
\mathsf{P}_x(t) = \begin{drop} 2 - 2t, \mathrm{if~~} t \in [0,1] \\
                        0,       \mathrm{otherwise}
           \end{drop} 
\]
\[
\mathsf{P}_y(t) = \begin{drop} 2t,\mathrm{if~~} t \in [0,1] \\
                        0,       \mathrm{otherwise}
           \end{drop} \\
\]
\[
\mathsf{P}_z(t) = \begin{drop} 1,\mathrm{if~~} t \in [0,1] \\
                        0,       \mathrm{otherwise}
           \end{drop} \\
\]
Evaluating the function $F_x$ at a point $a$ gives the probability
that clock $x$ is set to a value less than $a$, and if $a > b$, then
$F_x(a)-F_x(b)$ gives the probability that clock $x$ is set to a value
between $a$ and $b$, provided $a$ and $b$ are constants.  The same
calculation using the corresponding probability density function (pdf)
would
be $\int^a_b {\sf P}\!_x(x)dx$, which at first sight appears more
complicated.  The advantage is that these functions can be used to
calculate the probability that clock $x$ is set to a value less than
$y$, where $y$ is a random variable set according to the distribution
function 
$F_y$.  If, for example, we wished to calculate the probability of the
equivalence class in region 1 ($v_0(x) < v_0(y) < v_0(a)$, where
$v_0(a) = 1$) we would evaluate
$\int^y_0{\mathsf P}_x(x)dx$, to give us a function that returns the
probability that $v_0(x)$ is between 0 and $y$, multiply this by the
pdf 
${\mathsf P}_y(y)$, and integrate between zero and 
one:

\[\displaystyle
\int^1_0\int^y_0 {\mathsf P}_x(x)dx {\mathsf P}_y(y)dy
\]
which gives us the probability that $x$ will be less than $y$, where
$x$ and $y$ are random variables conforming to the distribution
functions $F_x$ and $F_y$.

We will now evaluate the probabilities of some of the arcs in the
example.  In the following, we will continue to subscript the clock
variables by the iteration number, in order to distinguish different
settings of the same clock.

In our example, to determine the probability on  arc $(0,2)$, where
the value to which clock $y$ is initially set (which we will refer to
as $y_0$) is less than the value to which clock $x$ was initially
set ($x_0$), ($y_0 < x_0$)  we perform the double integration
\[
\displaystyle \int_{0}^{1} \int_{0}^{x_0} 2y_0 dy_0 \,(2-2x_0) dx_0  
\]
which evaluates to $\frac{1}{6}$.

 Arc $(0,1)$  must have the value $1-\frac{1}{6} = \frac{5}{6}$, since
 it is the only other possibility, and
can be calculated as 
\[
\displaystyle \int_{0}^{1} \int_{x_0}^{1} 2y_0 dy_0 \,(2-2x_0) dx_0  
\]

These two arcs represent the setting of the clocks, and are therefore
instantaneous. 

From Region 2 the only  region which can be reached is the leaf node
region 5, and therefore the arc $(2,5)$ has probability 1.

Calculating probabilities on the paths through region 3 is more
complicated.  Consider arc $(3,6)$ first.  In fact, we must calculate
the probability of the path $(0,1,3,6)$ in its entirety rather than
determine separately the conditional probability of arc $(3,6)$.  We
do this as follows.

The clock setting information we know is: the first time the clocks
$x$ and $y$ are set, the initial value of $x$ is less than the initial
value of $y$ ($x_0<y_0$); and when $z_1$ is set, the sum of $x_0$ and
$z_1$ is less  than the initial value of clock $a$
($x_0 + z_1 < 1$).    These constraints are captured as the
combination of the integrals
$\int_{x_0}^1 {\sf P}_y(y_0) dy_0$
(to ensure that $x_0<y_0<1$),
$\int_{0}^{1-z_1} {\sf P}_x(x_0) dx_0$
(to ensure that $x_0+z_1<1$), and
$\int_{0}^1 {\sf P}_z(z_1) dz_1$
(since all constraints have been captured in the first two integrals.)

The combination is given as the first integral in Table~\ref{table:integrals} 
and  equals $\frac{3}{5}$.

\begin{table*}
\begin{center}
$
\displaystyle
\int_0^1 
        \int_{0}^{1-z_1}  
                        \int_{x_0}^1  {\sf P}_y(y_0) dy_0     
        {\sf P}_x(x_0)dx_0            
{\sf P}_z(z_1)dz_1
$
\\
\vspace{1cm}
$
\displaystyle
\int_0^1 
        \int_{1-z_1}^{1}  
                        \int_{x_0}^1  {\sf P}_y(y_0) dy_0     
        {\sf P}_x(x_0)dx_0            
{\sf P}_z(z_1)dz_1
$
\end{center}
\caption{The integrals} \label{table:integrals}
\end{table*}

The path $(0,1,3,7)$ differs only in the fact that $a_1$ ($=1-x_0$) is
less 
than $z_1$, and can be calculated as the second integral in
Table~\ref{table:integrals} which equals $\frac{7}{30}$.  
The only difference is that ${\sf P}_x(x_0)$ is integrated between $1-z_1$
and $1$.

At this stage in the algorithm, $\Sigma{\sf p} = \frac{1}{6}$
and $\Sigma{\sf f} = \frac{7}{30}$.  Since  $\Sigma{\sf f} > 1-0.9$ we
can deduce that the formula is false, and in
this case, there is no need to unfold further the node labelled
${\mathsf u}$.  

The accuracy with which we know the values of  $\Sigma{\sf p}$ and
$\Sigma{\sf u}$ will increase as the
probabilistic region tree is extended, and in some cases it may need
to be extended to infinity for perfect accuracy.  However, we can
achieve accuracy to within an arbitrary tolerance $\epsilon$ with a
finite probabilistic region tree.

 The major drawback of this algorithm is its complexity: with every
new unfolding of the probabilistic region tree not only does the
number of nodes to be considered increase, but also the number of
integrations required to determine the probability on a single node
increases exponentially.  It therefore becomes intractable after a few
iterations.  This is the issue we try to tackle with the second
algorithm.  Rather than integrate the probability density functions,
we discretise the ranges of the functions and work with the resulting
approximations.

\section{The Matrix algorithm}  \label{sec:alg-two}

In this section we present an overview of the second algorithm.  The
second algorithm takes a stochastic automaton {\em SA}, together
with a bounded until temporal logic formula TL, a time step parameter
$\delta$ and an adversary {\em pick}.  For convenience we will present
only the case where TL is of the form $[\phi_0\!\! \until_{\!\!\leq
c}\, \phi_1]>p$.  Minor modifications to the algorithm would allow any
of $\geq p$, $\leq p$ or $< p$.  We use the atomic propositions
$\phi_0$ and $\phi_1$ as part of the formula because anything more
complex can be reduced to these by standard model checking techniques.
Using $\leq c$ guarantees that the algorithm will terminate, although
we discuss the $\geq c$ case in Section~\ref{sec:greater-than}.

A single iteration of the algorithm will return one of
three results: ${\sf true}$, ${\sf false}$ or ${\sf undecided}$.  If
it returns ${\sf true}$, then the automaton models the formula.  If it
returns ${\sf false}$, then the automaton does not model the formula.
If it returns ${\sf undecided}$, then the algorithm was unable to
determine whether the automaton models the formula.  In this case, the
algorithm can be re-applied with a smaller value for the timestep
$\delta$.  The question of convergence to the correct answer as
$\delta$ tends to zero is discussed in
section~\ref{sec:proof}.  For the remainder of this section we
assume $\delta$ to be fixed.   

A stochastic automaton has a finite number of clocks each with a
probability distribution function (pdf).  For each state, the set of
clocks has an (arbitrary) order, and the algorithm makes use of this
ordering\footnote{However, the choice of ordering is arbitrary and
does not carry any meaning.  Any ordering will be sufficient.}. We
assume that each clock has non-zero lower and upper bounds on the
values to which it can be set. The first of these is a new constraint
and was not required for the first algorithm. This has been done so
that $\delta$ can be initially chosen to be less than the minimum of
all these lower bounds.

The algorithm works by creating a snapshot of the automaton at each
time point $n\delta$ ($n \in {\nat}$)\footnote{We will speak of the
time instants generated by $n\delta$ ($n \in \nat$) as time points.}
and extracting some global information about the probability of the
formula $[\phi_0\!\! \until_{\!\!\leq c}\, \phi_1]$ being satisfied at
this point.\footnote{We also require that $\exists n . n\delta = c$,
which ensures that one of the snapshots will be at exactly time $c$.}
To build the next snapshot, the algorithm picks out at each time point
$n\delta$ the transitions that the automaton is capable of during the
next interval of length $\delta$.  Because $\delta$ is less than the
minimum of all the clock lower bounds, a maximum of one transition per
path can occur in each interval.  Recording all possible states of the
automaton at each time point is therefore enough to record all the
possible transitions.

The algorithm stops when either enough information has been gathered
to determine the truth or falsity of the formula, or enough time has
passed so that $n\delta > c$, and allowing time to pass further will
make no difference to the information we already have.  In this case
the result {\sf undecided} is returned.  

\subsection{Data structures}

The principal data structures used by the algorithm are 
matrices.  For each state $s$ in the stochastic automaton we derive a
matrix for a given time $t$ (which is a rational number and calculated
as $n\delta$), denoted
$matrix(s,t)$, which is a record of the probabilities of the various
combinations of clock values in state $s$ at time $t$.

Each matrix $matrix(s,t)$ will have $\#\kappa(s)$ dimensions.  Each dimension is
associated with a particular clock, and the ordering of the dimensions
corresponds to the ordering of the clocks.  The dimension associated
with a clock $c$ will have $\lceil\frac{c_{max}}{\delta}\rceil$ entries,
where $c_{max}$ is the largest value to which the clock $c$ can be set,
and $\lceil \frac{c_{max}}{\delta} \rceil$ is the smallest integer
greater than or equal to $\frac{c_{max}}{\delta}$. For a clock $c_i$, we
will abbreviate $\lceil \frac{c_{i_{max}}}{\delta} \rceil$ by $N_i$.

The valuation function $v$ gives the value of a particular clock:
$v(c_i)$ is the value of clock $c_i$.  

Each entry in the matrix $matrix(s,t)$ is the probability that at time
point 
$t$, the automaton is in state $s$, and each clock is within a
particular time range. 
Thus, the value $matrix(s,t)[k_1 \ldots k_n]$ is the
probability that at time point $t$, the automaton is in state $s$,  and
 $v(c_i) \in (\delta(k_i - 1),\delta k_i]$ for each clock $c_i$. 

A further data structure we shall need is $live(t)$, which  is the set
of states ``live'' at time $t$ (i.e.\ their 
matrices at time $t$ contain at least one non-zero entry, and the formula
is still undecided).  In order to
get an accurate picture of the automaton at time $t+\delta$, we must take
into account all states live at time point $t$.

A $snapshot$ of the automaton at time $t$ is the set of all
matrices $matrix(s,t)$ where $s$ is in $live(t)$.  

Let $pr(c_i \in (\delta(k_i-1), \delta k_i])$ be the probability that
clock $c_i$ is initially set to a value in the range $(\delta(k_i-1),
\delta k_i]$.  Before the algorithm proper begins, we calculate all
these values from the clock probability distribution functions, which
are entered into the algorithm as part of the stochastic automaton.

\subsection{Variables}

The algorithm also uses a number of auxiliary variables.  

$prob(s,t)$ is  the probability of entering state $s$ 
during the time range  $(\delta(k-1), \delta k]$ (where $t=\delta
k$) and is defined
for 
states $s$ 
live at time  $\delta(k-1)$, and $s'$ live at time $\delta k$.  

$new\_states(s,t)$ is the set of states which can be reached from a
state $s$ during a time range $(\delta(k-1), \delta k]$. 

$total\_pass$ is a probability value.  It is incremented at each
iteration.  The iterations of the algorithm correspond to the 
time points, and $total\_pass$ records the probability of the
automaton having passed the formula at that time.  $total\_fail$ is
also a probability value; it records the probability of the automaton
having failed the formula as the algorithm progresses.

$error$  is an  upper bound on the possible errors of
$total\_pass$ and $total\_fail$.  After an iteration, we know that the
actual probability of the automaton having passed the formula is in
the range $[total\_pass,total\_pass+error]$, and similarly for
$total\_fail$.  

\subsection{Overview}

The second  algorithm is given in detail in
Appendix~\ref{sec:second-algorithm}.   We begin here with a pseudocode
description.   

\begin{tabbing}
...\=...\=...\=check \= formula against $s_0$ and $t=0$.\=...\=   \kill
build $matrix(s_0,0)$ \\
check formula against $s_0$ and $t=0$  \>\>\>\>\> $\rightarrow$ {\sf pass} \\
                      \>\>\>\>\> $\rightarrow$ {\sf fail} \\
\>\>\>\> $\downarrow$ {\sf undecided} \\
repeat \\
\> $t := t + \delta$ \\
\> forall locations $s$ in $live(t-\delta)$ \\
\>\> build $matrix(s,t)$ \>\>\>{\em (record possible new locations)} \\
\>\>\>\>\> {\em (increment probability of entering new locations)} \\
\>\>\>\>\> {\em (increment $error$)} \\
\>\> update $live(t)$ \\
\> forall locations $s'$ in $live(t)$ \\
\>\> check formula against location: \\
 \>\>\> if {\sf pass} then add probability to $total\_pass$ \\
\>\>\>  if {\sf fail} then add probability to $total\_fail$ \\
\>\>\>  if {\sf undecided} then update $matrix(s',t)$ \\
until \>\>\> (formula has passed, or \\
\>\>\> formula has failed, or \\
\>\>\> $t$ has reached the limit set by the formula) \\
set all locations {\sf undecided} at last iteration to false \\
if $total\_pass > formula probability$ then output {\sf pass}\\
elseif $total\_fail > 1- formula probability$ then output {\sf fail}\\
else  output {\sf undecided} \\
\end{tabbing}

We now present the formula for initially calculating matrices, then
describe the algorithm in overview, outlining the 
procedures involved. 

 If there are $n$ clocks in state $s$, then $matrix(s,t)$ is
calculated using the probability distribution functions of the clocks
in state $s$ as follows:

\begin{tabbing}
....\=....\=....\=....\=....\=....\=....\=....\=....\=....\=....\=....\=
\kill
\> $\forall 1 \leq k_1 \leq N_1$ \\
\>\>\> $\vdots$ \\
\> $\forall 1 \leq k_n \leq N_n \bullet matrix(s,t)[k_1\ldots k_n] :=
{\displaystyle \prod_{l=1}^{n}} pr(v(c_{l}) \in (\delta(k_l-1),\delta
k_l])$ \\ 
\end{tabbing}

The algorithm begins  by calculating $matrix(s_0,0)$,
where 
$s_0$ is the initial state of the stochastic automaton.

$live(0)$ will either be $\{s_0\}$ or the empty set, according
to whether the formula TL is made true or false by state $s_0$, or
whether we cannot yet decide.  This is determined as follows.  If
state $s_0$ models proposition $\phi_1$, then the formula TL is
immediately true and $live(0)$ is the empty set.  Otherwise, if $s_0$
models $\phi_0$ we cannot yet decide, and so $live(0)$ contains
$s_0$. If the state models neither proposition then the
formula TL is immediately false, and $live(0)$ is the empty set.
\vspace{12pt}

If the initial step does not determine whether the formula is
true or false, we perform a number of iterations.  Each iteration
builds the snapshot at time point $t+\delta$, based upon the snapshot
at time point 
$t$.   The sequence of snapshots build progressively more information
as to whether the stochastic automaton has passed or failed the
formula.  

In the case of a bounded until formula with a $\leq c$
subscript\footnote{i.e. $[\phi_0\!\! \until_{\!\!\leq {c}}\, \phi_1] > p$. See
Section~\ref{sec:greater-than} for a discussion of how $>\! c$ time bounds
are handled.}, the number
of iterations is finite (i.e.\ the algorithm always 
terminates) because the iterations terminate either when sufficient
information has 
been extracted to determine whether the formula passes or fails, or
after the $\frac{c}{\delta}$th iteration, since the formula cannot
become true after time $c$.

If the information at time $t$ is not enough to determine the truth
or falsity of the formula, we build the snapshot for time point $t+\delta$.
We now describe an individual iteration.  

An iteration consists of two sections.  In the first, we consider  all
of the states which are currently 
undecided.  These are all the states in $live(t)$.  For each state we
create the matrices at
time $t+\delta$, update $live(t+\delta)$ and calculate
$prob(s',t+\delta)$ for states $s'$ which can be reached in the
interval $(t,t+\delta]$.  In the second, we look at all states which
can be reached in the interval $(t,t+\delta]$, and  consider them with
respect to the temporal logic formula.  We then either update the global
probabilities, if the states cause the formula to pass or fail,
otherwise we update the respective matrices.

Note that in this algorithm a matrix is  updated at most twice.
Once within procedure $new\_time\_matrix$(refer to
Appendix~\ref{sec:second-algorithm}), if the state was live at
the previous time, and once within the procedure $new\_state\_matrix$,
if the state is reachable via a transition in the previous interval.  

\subsubsection{Creating and updating matrices}

We begin with some necessary notation.  Let us assume  $\delta$ is a
fixed rational number greater than zero.

\begin{Definition}
If $c_1,\ldots,c_n$ are the clocks on state $s$, a {\it 
  valuation}\footnote{We alter the definition of valuation slightly
  here for the second algorithm.} is
  the vector of results of the valuation function $v(c_i)$ from clocks 
  to ${\cal R}$ which gives 
the values of each of the $n$ clocks.  

Two valuations $v$ and $ v'$ are ($\delta -$) equivalent if 
\[
\forall c_i.\exists k_l \in \nat. v(c_i) \in (\delta(k_l -1),k_l] \land v'(c_i)
\in (\delta(k_l -1),k_l]
\]

A {\it valuation equivalence class} (or clock configuration) is  a
 maximal set of equivalent  valuations.  $\hfill\Box$

\end{Definition}

If $\delta$ is understood, we can abbreviate this configuration as
$(k_1,\ldots,k_n)$.   For a
state $s$ and a time point $t$, the probability $\prod_{l=1}^{n} pr(v(c_{l})
\in (\delta(k_l-1),\delta k_l])$ is an $(s,t)$-{\it clock configuration
  probability} (or just a clock configuration probability when $s$
and $t$ are understood).

%
%
%

\vspace{12pt}

There are two different procedures for updating a matrix.  The first
(encapsulated in the procedure $new\_time\_matrix$) corresponds to the
situation within the stochastic automaton where time passes, but the
state remains unchanged.  In this case we must shift the clock
configuration probabilities in the previous matrix down by one index
step (which corresponds to $\delta$ time passing) and add the result
to the matrix we are updating.

We also at this stage determine the new states which can be reached
from the current state during the $\delta$ time passing, and the
probability of entering these states.  We do this by looking at all
the clock configurations where at least one of the indices has the
value one.  If the clocks are set within such a configuration then we
know that at least one clock will expire during the ensuing $\delta$
timestep.

If only one index in the configuration has the value one then only one
clock can expire, and only one state can be entered from this clock
configuration, and so that state is added to the set of states which
can be entered from the current state at the current time.

If more than one index in the configuration has the value one, then
we simply do not go any further into the automaton and the
configuration probability is added to error. 

The second way to update a matrix corresponds to a transition from one
state to another within the automaton.  It is described in the
procedure $new\_state\_matrix$.  For each matrix entry we calculate
the clock configuration probability, multiply it by the probability of
moving into this state at this time, and add it to the matrix entry we
are updating.

\subsubsection{Termination of an iteration}

When the iteration terminates, it will output one of three results:
{\sf true}, {\sf false} or {\sf undecided}.  {\sf true} means that the
automaton models the temporal formula, i.e.\ $SA \models
[\phi_0\!\!  \until_{\!\!\leq c}\, \phi_1] > p$.  {\sf false} means that
$SA \not\models [\phi_0 \!\!\until_{ \!\!\leq c}\, \phi_1] > p$, and
{\sf undecided} means that the algorithm could not accumulate enough
information to decide whether or not the automaton modelled the
formula.

The algorithm makes the output decision based on the three global
variables $total\_pass$, $total\_fail$ and $error$.

$total\_pass$ is a lower bound on the probability that the stochastic
automaton models the formula, and $total\_fail$ is a lower bound
on the probability that the stochastic automaton does not model the
formula.  $error$ 
is the largest amount by which $total\_fail$ or $total\_pass$ may be
wrong.  In a sense, it records the size of the uncertainty introduced
by the choice of $\delta$. 

If neither of these situations holds then the errors introduced by the
algorithm are too large to determine an answer with this value of
$\delta$.  In this case, we can rerun the algorithm
with a smaller $\delta$, and in section~\ref{sec:proof} we show
that the sum of the errors tends to zero as $\delta$ tends to zero.
Note, however, that in the case where the probability that $SA$ models
$[\phi_0\!\! \until_{\!\!\leq c}\, \phi_1]$ is exactly $p$, we cannot
guarantee 
that there will be a $\delta$ small enough to allow the algorithm to
generate a {\sf true} or a {\sf false}.  This is the sort of
limitation that has to be accepted when working with generalised
distributions.

\section{Example}  \label{sec:example}

The second algorithm requires slightly more stringent restrictions on
the stochastic automaton than the first one, because the clock
distribution functions must have positive lower bounds, (as opposed to
the non-negative lower bounds required by the first). Therefore in
order to illustrate the second algorithm, we will use the 
automaton in Figure~\ref{fig:stoch-automaton}, but alter slightly each
of the clock distribution functions, by shifting each of them half a
time unit to become 
\[
\begin{array}{rcl}
F_x(t)  & = & 2t-t^2, \mathrm{if~~} t\in(\frac{1}{2},\frac{3}{2}] \\
        & = &      0, \mathrm{if~~}  t \leq \frac{1}{2} \\
        & = &    1, \mathrm{otherwise} \\
\end{array}
\]
\[
\begin{array}{rcl}
F_y(t) & = & t^2, \mathrm{if~~}  t\in(\frac{1}{2},\frac{3}{2}] \\
        & = &      0, \mathrm{if~~}  t \leq \frac{1}{2} \\
        & = &    1, \mathrm{otherwise} \\
\end{array}
\]
and
\[
\begin{array}{rcl}
F_z(t) & = & t, \mathrm{if~~}   t\in(\frac{1}{2},\frac{3}{2}] \\
        & = &      0, \mathrm{if~~}  t \leq \frac{1}{2} \\
        & = &    1, \mathrm{otherwise} \\
\end{array}
\]

In this section, we will consider the temporal formula $[(a_0 \lor
a_1) \!\!\until_{\!\!\leq \frac{3}{2}} a_2] > \frac{1}{2}$, where $s_i
\models a_i, i \in \{1,2,3\}$.

We now illustrate this algorithm by applying it to the
  example\footnote{The type of situation where this algorithm would do
  very badly is if one clock has a very small lower bound and all the
  rest have a very high lower bound.  This is accentuated if the first
  clock is hardly used.  It might even be that the state where the
  first clock is used is unreachable or has a very low probability of
  being reached.  Thus a criterion for the algorithm to work
  efficiently is that all pdf lower bounds are ``similar''.}. We set
   $\delta$ equal to $\frac{1}{2}$.

Sections A, B and C below correspond to the sections A,B and C in the
algorithm description in Appendix~\ref{sec:second-algorithm}.  Within
section 
C, line numbers correspond to the line numbers of the algorithm.

\subsubsection*{Section A} 

This section initialises all the variables to zero, and
  calculates all 
the probabilities of clocks falling in the ranges
  $(0,\delta],(\delta,2\delta]$ etc.  from the probability
distribution functions entered as part of the stochastic automaton. 

In our example, the probabilities that the clocks $x$, $y$ and $z$ are
in the ranges $(0,\delta], (\delta,2\delta]$ or $(2\delta,3\delta]$
are given by
\[
\begin{array}{rccc}
                 &    x        &      y      &     z  \\
(0,\delta]       &    0        &      0      &     0  \\
(\delta,2\delta] & \frac{3}{4} & \frac{1}{4} & \frac{1}{2} \\[3pt]
(2\delta,3\delta]& \frac{1}{4} & \frac{3}{4} & \frac{1}{2} \\
\end{array}
\]

These are easy to obtain from the clock probability distribution
functions.  Indeed, the ease of determining these probabilities is the
main benefit of this algorithm and contrasts with the intractable
manner in which the integrals explode in  the first algorithm.

\subsubsection*{Section B}

The initial state $s_0$ does not model $a_1$, but it does model the
proposition $a_0$, and so the procedure $init\_matrix$ is called.
This returns  $matrix(s_0,0)$ which is as follows
\[
\begin{array}{c|cccc}
y &   & & &\\
3 & 0 & \frac{3}{8} & \frac{1}{8} &  \\
2 & 0 & \frac{3}{8} & \frac{1}{8} &  \\
1 & 0 & 0 & 0 & \\
\hline 
  & 1 & 2 & 3 &  x
\end{array}
\]
and is easily derivable from the probabilities above.  The procedure
also sets $live(0)$ to $\{s_0\}$.   

If $N_x$ is the upper bound of $x$, and $N_y$ is the upper bound of
$y$, there will be $\lceil N_x \times \frac{1}{\delta}\rceil$ entries
on the $x$ axis, and $\lceil N_y \times \frac{1}{\delta}\rceil$
entries on the $y$ axis, so in this case (where $N_x=\frac{3}{2}$,
$N_y=\frac{3}{2}$ and $\delta = \frac{1}{2}$),  we get a $3\times 3$ matrix.

This matrix tells us e.g.\ that when the clocks in the initial state
are first set, the probability of clock $x$ being set within the range
$(\delta,2\delta]$ 
and clock $y$ being set within the range $(2\delta,3\delta]$ is $\frac{3}{8}$.
That is, for the clock configuration
$\trace{(\delta,2\delta],(2\delta,3\delta]}$, the clock
configuration probability is $\frac{3}{8}$.  

\subsubsection*{Section C}

We now enter the iterative part of the algorithm, where each iteration
corresponds to increasing the time by one time unit ($\delta$), and
the snapshot 
produced at the end of iteration $n$ corresponds to a view of the
automaton at time $n\delta$.  The three global probability
values\footnote{These are the probability values that are updated
  throughout the algorithm: $total\_pass, total\_fail$ and
  $error$.} are 
all still zero (lines 1-1a), so $ct$ (current time) becomes $\delta$.
Only the state $s_0$ is live at time zero, so $new\_time\_matrix$ is
called (line 6) for $matrix(s_0,\delta)$.  This returns a number of
parameters:  $matrix(s_0,\delta)$,
$new\_states(s_1,\delta),prob$ and $error$.

The procedure $new\_time\_matrix$ will return the  $matrix(s_0,\delta)$  as 

\[
\begin{array}{c|cccc}
y &   & & &\\
3 &  0 & 0 & 0 & \\
2 &  \frac{3}{8} & \frac{1}{8} & 0 & \\
1 &  \frac{3}{8} & \frac{1}{8} & 0 & \\
\hline 
  & 1 & 2 & 3 &  x
\end{array}
\]
where each clock has advanced one time unit from $matrix(s_0,0)$.  So,
at time $\delta$, the probability of clock $x$ being within the range
$(0,\delta]$ and clock $y$ being within the range $(\delta,2\delta]$
is $\frac{3}{8}$. 

The probability of staying in state $s_0$ for at least half a time
unit 
is 1; this follows from the fact that no clock can be set
to less than $\delta$ ($\frac{1}{2}$  time unit).  Thus
$prob(s_0,\delta) = 1$.   

None of the edge values (those with at least one clock in the range
$(0,\delta]$) of the previous time matrix ($matrix(s_0,0)$) is non-zero (so
there is no possibility of any clock reaching zero and causing a
transition to fire). The second half of the procedure (lines 10-23,
which would determine the new states reached from state $s_0$) is
therefore not executed and the global probability values
($total\_pass, total\_fail$ and $error$) are all
still zero. $new\_states(s_0,\delta)$ will be returned as $\{\}$,
since no new states can be reached at time $\delta$.  

The next step (lines 7-11 of section C) is to calculate the live
states at time $\delta$, and since $remain(s_0,\delta) = true$ (it is
possible to remain in state $s_0$ at time $\delta$) we
include $s_0$.

Since there are no states which can be reached from state $s_0$ in the
time interval $(0,\delta]$, lines 12-22 of section C are not executed.

All of the global probability values are still zero, (i.e.\ we don't
have enough information to decide the truth or falsity of the formula
at this stage, lines 1-1a of Section C), and $2\delta \leq 2$ (we have
more time in which to gain more information, lines 2-3 of Section C),
so we begin a second iteration.  

On the second iteration of the while loop, $ct$ is set to
$2\delta$. Only $s_0$ was live at the last iteration
($live(\delta)=\{s_0\}$), so at line 6 we call $new\_time\_matrix$ for
$matrix(s_0,2\delta)$.  

This again returns a number of parameters,
e.g.\ $matrix(s_0,2\delta)$ becomes
 
\[
\begin{array}{c|cccc}
y &   & & &\\
3 &  0 & 0 & 0 & \\
2 &  0 & 0 & 0 & \\
1 &  \frac{1}{8} & 0 & 0 & \\
\hline 
  & 1 & 2 & 3 &  x
\end{array}
\]
where the entry $matrix(s_0,2\delta)(1,1)$ is taken from the clock
configuration $(\delta,2\delta],(\delta,2\delta]$ in the previous time
matrix $matrix(s_0,\delta)$ and thus the probability of staying in
state $s_0$ in the interval $(\delta,2\delta]$ is $\frac{1}{8}$.
However this
is not the final version of $matrix(s_0,2\delta)$, because some of the
clock configurations lead to transitions which lead back to state
$s_0$.

All the other clock configurations ($(1,1)$, $(1,2)$ and $(2,1)$) in
$matrix(s_0,\delta)$ lead to transitions.   Lines 10-22 of procedure
$new\_time\_matrix$ 
are executed 
for each of these three configurations.  

For clock configuration $(1,1)$, clock $x$ is (arbitrarily) chosen to
fire, and we assume that the adversary $pick$ chooses the action
$conc$, leading to state $s_1$.  Line 13a of the procedure adds state
$s_1$ to $new\_states(s_0,2\delta)$, and $prob(s_1,2\delta)$ becomes
$\frac{3}{8}$ (line 14). Clock configuration $(1,1)$ is one where some
error may be introduced into the algorithm result.  Choosing clock $x$
and action $conc$ meant that we go to a state where the formula $TL$
can still be true, but choosing the other clock may not lead to such a
state.  We therefore allow for the possible error introduced here by
adding the clock configuration probability to $error$, which becomes
$\frac{3}{8}$.  Clock configurations $(1,2)$ and $(2,1)$ are dealt
with similarly, but $error$ remains constant.  

Now, the $new\_time\_matrix$ procedure is finished, and lines 7-11 of
Section C determine the value of $live(2\delta)$ which is
$\{s_0,s_1,s_2\}$, because at time $2\delta$ the automaton may be in
any state.

Lines 12-22 of Section C consider each new state that can be reached
in time interval $(\delta,2\delta]$.  State $s_0$ still allows the
temporal logic formula to be true, and so procedure
$new\_state\_matrix$ is called (line 17).  However,
$prob(s_0,2\delta)=0$,  and therefore
$matrix(s_0,2\delta)$ is not altered. 

State $s_1$ still allows the  temporal logic formula to become true
(line 13) and so procedure $new\_state\_matrix$ is called (line
17). The probability of entering state $s_1$ in this interval is
$\frac{6}{8}$, so $matrix(s_1,2\delta)$ is

\[
\begin{array}{c|cccc}
 & 0 & \frac{3}{8}  & \frac{3}{8} & \\
\hline 
  & 1 & 2 & 3 &  z
\end{array}
\]

In state $s_2$ the formula is true, and so $prob(s_2,2\delta)$
($\frac{1}{8}$) is added to $total\_pass$ (line 14).

In the final iteration, the global probability values become:
$total\_pass = \frac{1}{8}$, $total\_fail = \frac{3}{8}$ and $error =
\frac{4}{8}$.  The iterations stopped because the value of time became
too large --- not because the global probabilities contained enough
information to make a decision.  This means that $total\_pass$
($\frac{1}{8}$) is a maximum possible probability value of the formula
$[(a_0\lor a_1)\!\!  \until_{\!\!\leq \frac{3}{2}}\, a_1]$ (with any
clock ordering) and $total\_pass - error$ ($-\frac{3}{8}$) is a
minimum possible probability value.
 
 Thus, since we wish to determine whether the actual probability value
 is greater than $\frac{1}{2}$, the algorithm will output {\sf fail}.

 If we were interested in a similar formula with a probability value
 in the range $[0,\frac{1}{8}]$, we could reduce the size of $\delta$,
 and take snapshots (e.g.) every $\frac{1}{4}$ time unit.  This (for
 the reasons outlined in Section~\ref{sec:proof}) will reduce
 the size of the $error$ variable.  

\subsection{Unbounded until formulae} \label{sec:greater-than}

As just presented the second algorithm only handles until formulae of
the form
\[
[\phi_1 \!\!\until_{\!\!\leq c}\,\phi_2]\simeq p
\]
however a combination of the second and first algorithms yields a
method to verify unbounded until formulae, i.e. those  of the form 
\[
[\phi_1 \!\!\until_{\!\!> c}\,\phi_2]\simeq p
\]

The basic idea is to observe that verification of a formula such as
$\phi_1 \!\!\until_{\!\!> c}\,\phi_2$ can be split into a
conjunction of separate verifications

\begin{itemize}

\item[(a)]  Check that $\phi_1$ holds at all times until $c$ time units
have elapsed; and

\item[(b)] Check that there exists an $X>c$ such that $\phi_2$ holds at
time $X$, and that for all times strictly greater than $c$ and less
than $X$, $\phi_1$ holds.

\end{itemize}

Thus, we can model check formulae such as $[\phi_1 \!\!\until_{\!\!>
c}\,\phi_2]\simeq p$ in the following way.

\begin{itemize}

\item[(i)] Run (the obvious slight adaption of) the second algorithm
to check that (a) holds.  This will finish with a certain amount of
probability mass in the variable {\em total\_fail} and no probability
mass in {\em total\_pass}.  The reason for the latter is that pass
states can only be revealed once time has passed beyond $c$.  In
addition, $live(c)$ will indicate the locations that are still
undecided, i.e. from which we must explore further.

\item[(ii)] Run the first algorithm using $live(c)$ as the starting
locations and the initial timing regions determined from the remaining
matrices (this can be done in a straightforward manner).  However,
notice that running the first algorithm in this situation does not
incur the problems of intractability that it does in the general case.
Specifically, since the time bound on the until has been satisfied we
ostensibly only have an untimed until verification.  Consequently
probabilities can be assigned to nodes without requiring the global
clock to be taken into account and thus, they can be evaluated
``locally''.  Hence, the exponential explosion in the number of
integrals to be considered does not occur.

\end{itemize}

\section{Correctness and convergence}  
\label{sec:proof}

For a single run with fixed $\delta$, we wish to prove two things:
that the algorithm terminating with {\sf pass} implies that the
automaton models the formula, and that the algorithm terminating with
{\sf fail} implies that the automaton does not model the formula.

If the algorithm outputs {\sf pass} then the variable {\em
total\_pass} must be greater than $p$ (where $p$ is taken from the
temporal formula $[\phi_0 \!\!\until_{\!\!\leq c}\, \phi_1] > p$).
The only place where {\em total\_pass} gets incremented is line 14 of
section C (see full algorithm in
Appendix~\ref{sec:second-algorithm}). If the current state $q$ models
$\phi_1$ (and all previous states in the path model $\phi_0$) we add
the probability of entering the state $q$ at the current time
point. If the sum of these probabilities is greater than $p$ then the
algorithm outputs {\sf pass}.

We will consider the case when the algorithm outputs {\sf
pass}. Consider the initial state.  Note that for any clock
configuration, 
the probability of all paths which commence with the clocks being set
somewhere within this configuration is equal to the clock
configuration probability.  Furthermore, for an arbitrary state $s$
and time $c$ and configuration, the probability of all paths which go
through this configuration at this time is the probability of the
configuration multiplied by the probability of reaching that state at
that time.

The probability of reaching state $s$ at time $c$ is the second
parameter passed to the procedure {\it new\_state\_matrix}\footnote{In
fact, it is greater than or equal to this sum, because some routes
through the transition system may have passed or failed the formula
already, and therefore would be considered no further by the
algorithm.}.

If every valuation in a configuration corresponds to the same
automaton transition, and this transition is the final one in a path
which models the formula, then we add the clock configuration
probability (multiplied by the probability of reaching that state at
that time) to {\it total\_pass}.

This is the only way in which the algorithm adds to the variable {\it
total\_pass}.  Since the algorithm only outputs {\sf pass} if {\it
total\_pass} is greater than the formula probability $p$, it is clear
that the algorithm will only output {\sf pass} if the automaton models
the formula.

If more than one clock in the configuration is in the range
$(0,\delta]$ then more than one of the clocks will have reached time 0
in the interval we are considering, and so the clock configuration
probability is added to {\it error}  (line 12 of procedure {\it
new\_time\_matrix}). 

A similar argument applies in the case where the algorithm outputs
{\sf fail}.

Therefore the algorithm is sound in the sense that if we are given  a
definitive answer, this answer is correct.  There remains, of course,
the question of convergence to the correct answer, and the following
theorem summarises the situation. 

\begin{Theorem} For every automaton $SA$ and propositions $\phi_0$ and
$\phi_1$ it is the case that if $SA$ models $[\phi_0
\!\!\until_{\!\!\leq c}\, \phi_1]$ with probability $p$, then for any
error $e$ greater than zero, there is a timestep $\delta$ greater than
zero such that for the formula $[\phi_0 \!\!\until_{\!\!\leq c}\,
\phi_1]\!>\!  q$, the algorithm will only return {\sf undecided} if $q
\in [p\!-\!e,p\!+\!e]$.

\end{Theorem}

 First note that $n$ independent single
variable continuous probability distribution functions
$f_1\ldots f_n$ can always be combined to give a single $n$ variable
probability distribution function which is continuous in all
dimensions:  $f(x_1\ldots x_n) = f_1(x_1) \times \cdots \times
f_n(x_n)$.  

For convenience, consider a location with two outgoing transitions and
two clocks $x$ and $y$ with distribution functions $f_x$ and $f_y$.
Because $f_x$ and $f_y$ are both continuous, if we set $f(x,y) =
f_x(x) \times f_y(y)$ we can (by the note above) say that
\[
\forall \epsilon > 0. \exists \delta > 0. f(x,x+\delta) -
f(x,x-\delta) < \epsilon
\]

We will show that for any desired size of error we can choose a
suitably small timestep.

\begin{figure}
\begin{center}
  \ \psfig{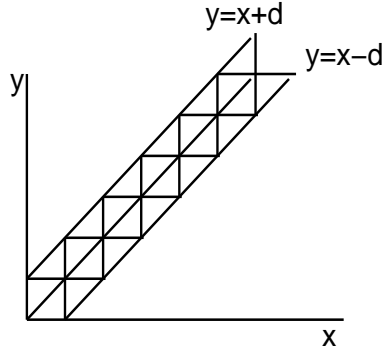}

\caption{Upper bound on error with clocks $x$ and $y$.}
\label{fig:proof.pic}
\end{center}
\end{figure}

Now, $\int_0^m f(x,x+\delta) - f(x,x-\delta) dx$ \footnote{ $m =
{min\{x_{max},y_{max}\}}$, where $x_{max}$ is the 
largest value to which  clock $x$ can be set.} (the probability of the clock
valuation falling between the two 45 degree lines in
Figure~\ref{fig:proof.pic}) is greater than the 
sum of all contributions to the error variables (represented by the
squares in the figure).  Since the number of locations in the stochastic
automaton is finite (say $N_s$) and (for bounded until formulas with less
than subscripts) the maximum number of visits to any location is finite
(say $N_v$) for any desired error $e$ we must ensure that, for every
location, for the multivariate function associated with that location, we
choose $\epsilon$ such that $\epsilon < \frac{e}{N_s \times N_v}$.  If
the timestep is set to the smallest $\delta$ necessary to ensure that
every location provides errors less than $\frac{e}{N_s \times N_v}$, then
total error provided by one location (over all time) will be less than
$\frac{e}{N_s}$ and the total error provided by all locations will be
less than $e$.

\section{Complexity measures}  \label{sec:complexity}

\subsection{Time complexity}

The time complexity of the algorithm discussed in
Section~\ref{sec:alg-two} depends on a number of factors, namely
$\delta$, $t$, $n_1$, $n_2$ and $\mid {\cal S} \mid$.  The explanation
of these parameters is as follows:

\begin{itemize}
\item $t$ is the value of time given in the time-bounded until
formula: $[a \!\!\until_{\!\!\!\leq t} b] \sim p$;
\item $\delta$ is the chosen timestep;
\item $\mid {\cal S} \mid$ is the number of states in the automaton;
\item  $n_1$ is the largest number of clocks in  a single state and
\item $n_2$ is the largest (positive  finite) upper bound of all the
clocks. 
\end{itemize}

 An upper bound on the number of matrices which need to be built in a
 single iteration is $\mid {\cal S} \mid$, where ${\cal S}$ is the set
 of all states in the automaton.

To calculate the time complexity we also need to calculate the size of
the largest matrix.  Each matrix is multi-dimensional, and
$\frac{n_2}{\delta}$ will be the maximum number of entries over all
matrices and all dimensions.  For example, in the example in
Section~\ref{sec:example} all the matrices had 2 dimensions and the
maximum number of entries in any dimension was 3 since $\delta =
\frac{1}{2}$ and $n_2 = \frac{3}{2}$.

An upper bound on the size of the largest matrix will therefore be
the   number 
of elements in the largest dimension, 
raised to the power of the largest number of clocks on a single
state.   

The time complexity is thus bounded by the time taken to update all
the possible matrices in each iteration of the while loop in the
algorithm, multiplied by the maximum number of iterations the
algorithm will perform in the worst case.  This latter value is
$\frac{t}{\delta}$, therefore the time complexity is  
\[
\frac{t}{\delta} \times
(\frac{n_2}{\delta})^{n_1} \times \mid {\cal S} \mid
\]

Although this is exponential, the exponent $n_1$ is something which
should in general be fairly small ($\leq 3$) because we only allow
clocks to be used from the state in which they are set.

 In fact, the algorithm could be optimised to provide a
better time complexity, by limiting the size of the matrices to
$\min(\frac{t}{\delta},\frac{n_2}{\delta})$ since there is no need
to consider the operation of the clock beyond the limit set by the
time bound on the temporal formula.  The size of the largest matrix
would 
therefore be  less than
$(\min(\frac{t}{\delta},\frac{n_2}{\delta}))^{n_1}$, 
where $n_1$ is the largest number of clocks in  a single state.

An upper bound on the time complexity would therefore be
\[
\frac{t}{\delta} \times
(min(\frac{t}{\delta},\frac{n_2}{\delta}))^{n_1} \times \mid {\cal S} \mid
\]

The time complexity  also relies heavily on $\delta$, and the bigger the $\delta$ the
lower the time complexity.  To see the relationship with $\delta$,
note that  the upper bound can be rewritten as 

\[
\left(\frac{1}{\delta}\right)^{n_1+1} \times t \times 
(n_2)^{n_1} \times \mid {\cal S} \mid
\]

\subsection{Space complexity}

An upper bound on the space complexity will be proportional to the
product of the size of the biggest matrix and the largest number of
matrices which need to be stored at one time.   The size of the
largest matrix is less than
$(\frac{n_2}{\delta})^{n_1}$, (from time
complexity calculations) and the largest number of matrices which need
to be stored at any one time is twice the number of states in the
automaton, $2 \;\times \mid {\cal S} \mid$.  The upper bound on space
complexity is therefore 
\[
2 \times (\frac{n_2}{\delta})^{n_1} \times \mid
{\cal S} \mid
\]

\section{Conclusions and further work} \label{sec:conclusions}

In this paper we have presented two algorithms for model checking
bounded  until formulae against stochastic automata.  Both of these
algorithms allow systems to be described using continuous probability 
distributions, and we believe that this represents an important
advance.  

 The principal
advantage of the first algorithm is its generality:  the clocks may be set
according to any function, providing the corresponding probability
density function is integrable.  The major drawback of the algorithm
is its complexity:  with every new unfolding of the probabilistic
region tree not only does the number of nodes to be considered
increase, but also the number of integrations required to determine
the probability on a single node increases exponentially.   

The principal advantage of the second algorithm is its efficiency: the
discretisation of the probability functions means that the
calculations required are considerably simpler.  A limitation in
comparison to the first algorithm is that the probability distributions
must have a finite lower bound.

In addition, an advantage of both the algorithms is that, since the
``complete'' model is at no point generated, the state space explosion
(which typically hinders model checking) is contained.  In particular,
all data structures apart from those which reflect undecided nodes
(i.e. {\sf u} 
labelled regions in the first algorithm and {\em live} locations in
the second algorithm) can be deleted.   In this sense the algorithms
yield a form of on-the-fly exploration -- only keeping information
about the ``leaves'' of the exploration tree.

  Further work on the second algorithm will include relaxing the
restrictions imposed on the stochastic automata, particularly the
ability to set and use clocks anywhere in the automaton.  Being able
to do this would allow parallel composition.

It would also be good to increase the expressiveness of the logic,
allowing nested untils or ``greater than'' queries, and to extend the
model checking algorithm itself to allow queries such as ``what is the
probability of $[\phi_0 \!\!\until_{\!\!\leq c}\, \phi_1]$?'' and receive
a probability value for an 
answer.

{\bf Acknowledgements: }
The research presented here is supported by the
UK Engineering and Physical Sciences Research Council under grant number
GR/L95878 (A Specification Architecture for the Validation of Real-time
and Stochastic Quality of Service).
Thanks are due to co-workers on this project for their input into this
work: 
 Gordon and Lynne Blair from Lancaster
University. Also to Pedro D'Argenio, Joost-Pieter Katoen and Holger
Hermanns.   In
particular Pedro's observations on clock equivalences for stochastic
automata have greatly influenced our approach.  

\bibliography{spa} 

\bibliographystyle{esub2acm.bst}

\setlength{\parskip}{1ex}
\setlength{\parindent}{0ex}

\appendix

\section{Semantics} \label{sec:stoch-semantics}

\subsection{Probabilistic Transition Systems}  \label{decoration} \label{pts}

The definition of the semantics of stochastic automata is given in
terms of probabilistic transition systems. The definition of
probabilistic transition systems is reproduced
from~\cite{D'Argenio-Katoen-Brinksma-1998}.

$\nat$ is the set of non-negative integers.  $\real$ is the set of
real numbers, and $\real_{\geq0}$ the set of non-negative
reals. For $n \in \nat$, let $\real^n$ denote the $n$th cartesian
product of $\real$.  $\real^0 \defeq \{\emptyset\}$. 

A {\it probability space} is a structure $(\Omega,{\cal F},P)$ where
$\Omega$ is a {\it sample space}, ${\cal F}$ is a $\sigma${\it
  -algebra} on $\Omega$ and $P$ is a {\it probability measure} on
${\cal F}$.  In this work,  as in~\cite{D'Argenio-Katoen-Brinksma-1998}, we
consider only probability spaces isomorphic to some Borel space
defined in a real hyperspace, whose coordinates come from independent
random variables.  We denote by ${\cal R}(F_1,\ldots F_n)$ the
probability space $(\real^n,{\cal B}(\real^n),P_n)$ where ${\cal
  B}(\real^n)$ is the Borel algebra on $\real^n$ and $P_n$ is the
probability measure obtained from $F_1 \ldots F_n$, a given family of
distribution functions. See~\cite{Shiryayev-1984} for details.

Let ${\cal P} = (\Omega, {\cal F}, P)$ be a probability space.  Let
${\cal D}:\Omega \rightarrow \Omega'$ be a bijection.  We lift ${\cal
  D}$ to subsets of $\Omega$: ${\cal D}(A) \defeq \{{\cal D}(a)\mid a
\in A\}$ and define ${\cal F}' \defeq \{{\cal D}(A) \mid A \in {\cal
  F}\}$.  Now, it is clear that ${\cal D}({\cal P}) \defeq
(\Omega',{\cal F}',P \circ {\cal D}^{-1})$ is also a probability
space. Since ${\cal D}({\cal P})$ is basically the same probability
space as ${\cal P}$, we say that ${\cal D}$ is a {\it decoration} and
we refer to ${\cal D}({\cal P})$ as {\it the decoration of ${\cal P}$
  according to ${\cal D}$.} This is used when we come to give a
semantics to stochastic automata.

\begin{Definition}  Let $P(H)$ denote the set of probability spaces
$(\Omega, {\cal F}, P)$   such that $\Omega \subseteq H$.  A {\it
  probabilistic transition system}  is a structure ${\cal T} =
  (\Sigma,\Sigma', \sigma_0,{\cal L},T,\longrightarrow)$ where 
\begin{enumerate}
\item $\Sigma$ and $\Sigma'$ are two disjoint
 sets of {\it states}, with the {\it initial state} $\sigma_0 \in
 \Sigma$.  States in $\Sigma$ are called {\it probabilistic states}
 and states in $\Sigma'$ are called {\it non-deterministic} states. 
\item ${\cal L}$ is a set of {\it labels}.  
\item $T: \Sigma \rightarrow P(\Sigma')$ is the {\it probabilistic
    transition relation}. 
\item $\longrightarrow \subseteq \Sigma' \times  {\cal L} \times 
  \Sigma$ is the {\it labelled (or non-deterministic) transition
    relation}.  We use  $\sigma' \stackrel{l}{\longrightarrow} \sigma$
  to denote $\trace{\sigma',l,\sigma} \in \longrightarrow$,
  $\sigma' \not\stackrel{l}{\longrightarrow}$ for
  $\lnot\exists\sigma.\sigma'\stackrel{l}{\longrightarrow}\sigma$ and
  $\sigma' \longrightarrow \sigma$ for $ \exists l. \sigma'
  \stackrel{l}{\longrightarrow} \sigma$.  $\hfill\Box$ 
\end{enumerate}                        
\end{Definition}

Since we are interested in timed systems, we set ${\cal L} = \mathbf{A}
\times  \real_{\geq0}$, where \textbf{A} is a set of action names.  A
timed action transition will be described as $a(d)$, which indicates
that the action $a$ occurs exactly $d$ time units after the system has
been idling.

\begin{Definition}{\upshape
  A {\em valuation} is a function $v:{\cal C} \rightarrow
  \real_{\geq0} \union \{\perp\}$ such that $v(x) \leq x_{max}$, where
  $x_{max}$ is the maximum 
  value to which clock $x$ can be set.  The set of all valuations is
  ${\cal V}$.  If $d \in \real_{\geq0}$, $v-d$ is defined by $\forall
  x \in {\cal C}. (v-d)(x) \defeq v(x)-d$.  We assume the set of
  clocks is ordered so, if $C \subseteq {\cal C}$, we can write
  $\stackrel{\rightarrow}{C}$ for the ordered form of $C$ and
  $\stackrel{\rightarrow}{C}\!\!(i)$ for the $i$-th element.  Let $C
  \subseteq {\cal C}$, $n = \#C$, and $\stackrel{\rightarrow}{D} \in
  \real^n$.  We define
  $v[\stackrel{\rightarrow}{C}\mapsleft\stackrel{\rightarrow}{D}]$ by
\begin{eqnarray*}
v[\stackrel{\rightarrow}{C}\mapsleft\stackrel{\rightarrow}{D}](x) 
& \defeq & \left\{
   \begin{array}{ll}
   \stackrel{\rightarrow}{D}(i) & {\mathrm{ if~~}} x =
   \stackrel{\rightarrow}{C}(i), {\mathrm{for~~ some~~}} i \in
   \{1,\ldots,n\} \\
   \perp & {\mathrm{otherwise}} \\
   \end{array}
\right.
\end{eqnarray*} $\hfill\Box$
}\end{Definition}
This definition will be used when we explain how clock values change
as states change.  It differs from the definition given
in~\cite{D'Argenio-Katoen-Brinksma-1998} because there clocks not in
the set ${\cal C}$ maintain their values through this operation.  This
is because in~\cite{D'Argenio-Katoen-Brinksma-1998} clocks may be used
to trigger actions in any state, not just the state in which they are
set.  In this work, however, in order to simplify the model checking, we
insist that clocks are only used in the states in which they are set,
and therefore there is no need to remember their value once the state
has been exited.

  The main obstacle now in constructing the
probabilistic transition system semantics is in showing how the
clock probability functions are used to construct the probability
spaces.  We do this by defining a decoration function, discussed in
Section~\ref{decoration}.  

Let $SA = ({\cal S},s_0,{\cal C},{\bf A},\blackright,\kappa,F)$ be
a stochastic automaton. Let $s$ be a location in ${\cal S}$ and $n =
\#\kappa(s)$.  Let $v$ be a valuation in ${\cal V}$. Let ${\cal V'} =
\{v[\stackrel{\rightarrow}{\kappa(s)} \mapsleft
\stackrel{\rightarrow}{D}] \mid \stackrel{\rightarrow}{D} \in
\real^n\} \subseteq {\cal V}$.  We define the decoration function ${\cal
  D}^s_v : \real^n 
\rightarrow \{s\} \times  {\cal V'} \times  \{1\}$ by ${\cal
  D}^s_v(\stackrel{\rightarrow}{D}) \defeq
(s,v[\stackrel{\rightarrow}{\kappa(s)} \mapsleft
\stackrel{\rightarrow}{D}],1)$.  Notice that ${\cal D}^s_v$ is a
bijection.  In the next definition, we use the probability space ${\cal
    R}(F_{x_1},\ldots,F_{x_n})$ decorated according to some ${\cal
    D}^s_v$. 

\begin{Definition}
  Let $SA = ({\cal S},s_0,{\cal C},{\bf A},\blackright,\kappa,F)$
  be a stochastic automaton.  The actual behaviour of $SA$  is given
  by the PTS $I(SA) \defeq (({\cal S} \times  
  {\cal V} \times  \{0\}), ({\cal S} \times  {\cal V} \times  \{1\}),
  (s_0,{\bf 0},0), \mathbf{A} \times  \real_{\geq 0}, T,
  \longrightarrow)$, where in the initial valuation {\bf 0} clock
  $a$ is set to some natural number (chosen according to the PRTL
  function, see Section~\ref{sec:PRTL}), and each other clock is
  undefined.  $T$ and $\longrightarrow$ are defined as follows:
\[
\begin{array}{rl}\frac{\stackrel{\longrightarrow}{\kappa(s)} =
    \{x_1,\ldots,x_n\}}{T(s,v,0) = {\cal D}^s_v({\cal
    R}(F_{x_1},\ldots,F_{x_n}))} & {\bf Prob} \end{array}
\]
\[
\begin{array}{rl}\frac{\begin{array}{c}s
    \!\stackrel{a,\{x\}}{\blackright}\!s' \land d \!\in  
      \!\real_{\geq 0} \land (v - d)(x) \leq 0 \\ \forall d' \in
      [0.d). \forall s'.s \stackrel{b,\{y\}}{\blackright}s'.(v - d')(y) >
      0\end{array}}{(s,v,1) \stackrel{a(d)}{\longrightarrow}
      (s',(v-d),0)} & {\bf Act} \end{array}
\]                  $\hfill\Box$
\end{Definition}

Within a stochastic automaton, two forms of uncertainty may arise.
One is the probabilistic uncertainty associated with the
clock-setting.  Although we 
know which clocks are to be set, the choice of values for these clocks
is probabilistic.  This is where the stochastic element of the
model arises, and is defined by rule {\bf Prob}.    The other is the
nondeterministic uncertainty that arises if two actions 
are simultaneously able to be performed, and is defined using the rule
{\bf Act}.  This nondeterminism is resolved using an
adversary (Definition~\ref{def:adversary}).

Definition of a PTS-path:

\begin{Definition}{\upshape
A {\em PTS-path}  is a {\em finite} or {\em infinite} sequence of
states  
\[
\trace{\sigma_0,\sigma'_0,\sigma_1,\sigma'_1,\ldots}
\]
where, $\sigma_0$ is the initial state, for each $\sigma'_i$, there
exists a probability space $(S, {\cal F},P)$ such that $T(\sigma_i)
=(S,{\cal F},P)$, $\sigma'_i \in S$ and $\sigma'_i \longrightarrow
\sigma_{i+1}$.  }    $\hfill\Box$
\end{Definition}

\begin{Definition}{\upshape
An {\em SA-path} is a finite or infinite sequence
\[
\trace{(s_0, v_0), (s_0, v'_0), (s_1,v_1), (s_1,v'_1), \ldots,
  (s_n,v_n), (s_n,v'_n),\ldots}
\]
such that 
\begin{itemize}
\item $v_0$ means no clocks are set. 
\item $v'_i \in {\cal R}(F_{x_1},\ldots,F_{x_n})$ where
  $T(s_i,v_i,0) = {\cal D}^s_v({\cal R}(F_{x_1},\ldots,F_{x_n}))$.
  Each valuation $v'_i$ is a possible result of
  the clock setting functions. 
\item $(s_i,v'_i,1) \stackrel{a(d)}{\longrightarrow}
  (s_{i+1},v_{i+1},0)$ for some $d$.    Timed action
  transitions must be allowed by the SA.
\item Finite paths end on a probabilistic state. 
\end{itemize} $\hfill\Box$
}\end{Definition}

An SA-path is like a run of the SA expanded with clock values.

\begin{Definition}{\upshape
\label{def:adversary}
  An {\em adversary} of an SA is a function mapping sequences of
  states to states 
\[
adv:  <s_0,s_1,\ldots,s_n> \longrightarrow s_{n+1}
\]
such that $<s_0,s_1,\ldots,s_n,s_{n+1}>$ is a run of the SA. $\hfill\Box$
}\end{Definition}
Note that adversaries do not make any reference to time. 

With an adversary, an SA becomes deterministic.  The corresponding PTS
contains no nondeterminism either.

 If 
\[
\sigma = \trace{(s_0, {\bf 0}), (s_0, v'_0), (s_1,v_1), (s_1,v'_1), \ldots,
  (s_k,v_k), (s_k,v'_k)}
\]
 is
  a finite  SA-path, then  $\sigma[i] = s_i$ and $\sigma(x)$ is the state at
  time $x$.  

${\cal R}(F_{x_1},\ldots,F_{x_n})$  is the Borel space $(\real^n,
{\cal B}(\real^n),P_n)$ where $P_n$ is the unique probability measure
obtained from ${\cal R}(F_{x_1},\ldots,F_{x_n})$.

Now, for all $j<k$, set $A_j$ to be the maximal set of valuations
equivalent to $v_j$ which lead to state $s_{j+1}$.  

%
%
%

Let
\[
C(s_0,A_0,s_1,\ldots,s_{k-1},A_{k-1},s_k)
\]
denote the {\em cylinder set} which contains all paths starting at
$s_0$ and going through all states $s_j (j\leq k)$ and valuation sets
$A_j (j\leq k)$.


The probability measure $Pr$ on ${\cal
F}(Path(s_0))$\footnote{$Path(s_0)$ is all paths possible from $s_0$,
and ${\cal F}(Path(s_0))$ is the smallest $\sigma-$algebra on $Path(s_0)$.}
is identified by induction on $k$ by $Pr(C(s_0))=1$ and for $k\geq0$:
\[
Pr(C(s_0,A_0,\ldots,A_{k},s_{k+1}) =
Pr(C(s_0,A_0,\ldots,A_{k-1},s_{k}))\cdot P(A_k)
\]

where $P(A_k)$ is the probability of the set $A_k$, and is taken from
the relevant Borel space.


%
%
%

\subsection{PRTL Semantics} 

In this section, we introduce the semantics for the temporal logic
PRTL. 

To facilitate model checking, we use Probabilistic Transition Systems
as a semantic model for the definition of PRTL.  But in order to do this we
must resolve two problems.  The first is that PRTL is a {\em
  real-time} logic --- it enables reference to specific instants in
time --- and the abstract definition of
PTSs~\cite{D'Argenio-Katoen-Brinksma-1998} does not contain reference
to time.  This is easily solved --- we simply use the PTS generated by
a Stochastic Automaton.  This contains much more detailed state
information, in particular, the values of clocks.

The second problem is that the PTS contains nondeterministic
information, and this nondeterminism must be resolved before we can
use the PTS to assign a semantics to our logic.  We do this using 
  adversaries.

%
%
%

%
%
%

Recall the syntax of PRTL:

\[
\begin{array}{c}
\psi ::= \mathsf{tt} \mid \mathsf{ap} \mid \lnot \psi \mid \psi_1
\land \psi_2 \mid [\phi_1 \!\! \until_{\sim c}\, \phi_2] \simeq p 
\\
\phi ::= \mathsf{tt} \mid \mathsf{ap} \mid \lnot \phi \mid \phi_1
\land \phi_2
\end{array}
\]

where $c \in \nat$, $a$ is an atomic proposition,  $p \in [0,1]$ is a
probability value and $\sim, \simeq \in\{<,>,\leq,\geq\}$.

The {\em path formulae} $\psi$ can only be used at the outermost level
--- they cannot be nested.   This is because the model checking
algorithms  only evaluate path formulae from the initial
state.

\begin{Definition}{\upshape
  If $SA = ({\cal S},s_0,{\cal C},{\bf A},{\blackright},\kappa,F)$ is
  a Stochastic Automaton and $PTS = (\Sigma,\Sigma', \sigma_0,{\cal
    L},T,\longrightarrow)$ is the resulting Probabilistic Transition
  System, then $\Sigma (= \Sigma') \subseteq {\cal S} \times  {\cal V}$, ${\cal
    L} \subseteq A \times  \real_{\geq 0}$ and $\sigma_0 = (s_0,{\bf
    0})$.  We must
  also introduce a function $\xi$ which maps SA locations to the
  logical propositions true in that location. $\hfill\Box$
}\end{Definition}

We only need to use the probabilistic states to define the logic,
since once a probabilistic state has been entered the behaviour of the
automaton is completely determined until the first clock expires.

The simple formulae $\phi$ are defined in the conventional way for
each probabilistic region $\sigma'$, but the until formulae $\psi$ are
defined only for the initial region $\sigma_0$.  The model checking
algorithm does not yet allow path formulae to be established for an
arbitrary region.  

\begin{itemize}

\item[$\bullet$] $s \models \mathsf{tt}$

\item[$\bullet$] $s \models a$, provided $a \in
  \xi(s)$

\item[$\bullet$] $s \models \phi_1 \land
  \phi_2$, provided $s \models \phi_1$
  and $s \models \phi_2$
\item[$\bullet$] $s \models \lnot  \phi$,
  provided $s \not\models \phi$

\end{itemize}

\vspace{12pt}

If $\sigma$ is an SA-path, and $\psi$ a path formula then
\begin{itemize}
\item[$\bullet$] $\sigma \models [\phi_1 \!\! \until \phi_2]$ iff $\exists k\geq
0 . (\sigma[k] \models \phi_2 \land \forall 0 \leq i \leq k. \sigma[i]
\models \psi_1)$
\item[$\bullet$] $\sigma \models [\phi_1 \!\! \until_{\!\sim t} \phi_2]$ iff
$\exists x\sim t . (\sigma(x) \models \phi_2 \land \forall y \in
[0,x). \sigma(y) \models \psi_1)$
\end{itemize}
and
\begin{itemize}
\item[$\bullet$] $PTS \models [\phi_1 \!\! \until_{\!\sim t}
\phi_2] \simeq p$ iff $Prob(s_0,\phi_1 \!\! \until_{\!\sim t} \phi_2)
\simeq p$ where $Prob(s_0,\psi) \defeq Pr\{\rho \in Path(s_0) \mid
\rho \models \psi \}$
\end{itemize}


Therefore, the Probabilistic Transition System $PTS$ models the
PRTL $[\phi_1 \!\! \until_{\!\sim t}
\phi_2] \simeq p$ provided $Prob(s_0,\phi_1 \!\! \until_{\!\sim
t} \phi_2) \simeq p$.

\section{First Algorithm}
\label{sec:first-algorithm} 

Here, we give the definition of the first model checking algorithm for
bounded until formulae.  We will consider a PRTL formula of the form
$[\phi_1 \until_{< c}\, \phi_2] > p$.  ``less than $p$'' queries may be
handled in a similar way.  

Assume an adversary {\em Adv}, and that each SA location is mapped to
either $\phi_1$ or $\lnot \phi_1$ and to either $\phi_2$ or $\lnot
\phi_2$.  Note that the algorithm can easily be extended to the more
general case where locations contain set of atomic propositions.

Add the (new) clock $a$ to the set of all clocks. 

Construct the PRG node  $(s_0,{\bf 0}_c)$.    

Set $s = s_0$.  

If $s \not\models \phi_1$ then stop with no, else 

REPEAT
\setlength{\parskip}{0ex}
\setlength{\parindent}{2ex}

For each  possible valuation equivalence class $[v_i]$ from
$\kappa(s)\bigcup \{a\}$, form the node  $(s,[v_i])$.

For each new node $(s,[v_i])$ choose a  subsequent  non-deterministic 
node $(s_j,\bot)$ according to the adversary {\em Adv}. 

\setlength{\parindent}{4ex}
   For each new
    non-deterministic node $(s_j,\bot)$ 

\setlength{\parindent}{6ex}
    label `{\sf p}' if $s_j \models \phi_2$ and $v(a) > 0$.   

    label `{\sf f}' if $s_j \not\models \phi_1$ or $s_j \not\models
    \phi_2$ or $v(a) \leq 0$.

    label `{\sf u}' otherwise

\setlength{\parindent}{4ex}

\setlength{\parindent}{2ex}

For each node labelled with either `{\sf p}' or `{\sf f}', calculate 
  the probability of the corresponding path.

If $\Sigma_{\sf p} pr(s,[v]) > p$ then stop with yes.

If $\Sigma_{\sf f} pr(s,[v]) > 1-p$ then stop with no.

Otherwise, repeat for each node labelled `{\sf u}'.

\section{Second algorithm}  \label{sec:second-algorithm}

In this section we present a detailed description of the algorithm.
It is divided into Section A (which initialises variables), Section B
(the initial part of the algorithm) and Section C (the iterative
part).  Procedures used are described at the end.  

The lines of code are prefaced with numbers, and the comments are
delimited with double stars.  

\begin{tabbing}
....\=....\=....\=....\=....\=....\=....\=....\=....\=....\=....\=....\= \kill
\> \bc{ Section A}\ec\\
 $ Model\_check(SA,Formula,\delta,pick)$    \\
\> \bc{ note that the function $pick$ is the adversary, used in procedure $new\_time\_matrix$.}\ec\\
\> \bc{ We are assuming a TL formula of the form $[a_0 \until_{\leq t} a_1] \geq p$.  }\ec\\
\> \bc{ The $\geq p$ could easily be changed; the $\leq t$ is hardwired into the algorithm. }\ec\\
\> \bc{ }\ec\\
\> \bc{ We begin by initialising variables.}\ec\\
\> \begin{com}{ $ct$: (integer)  current\_time}\end{com}\\
\> $ct := 0 $    \\
\> \bc{ $total\_pass$ and $total\_fail$ are reals in $[0,1]$.  }\ec\\
\> \bc{  At any point in the algorithm, $total\_pass$ is the accumulated }\ec\\ 
\> \bc{ probability of all the passed paths and $total\_fail$ is the accumulated }\ec\\
\> \bc{  probability of all the failed paths. We initialise them both to zero.}\ec\\
\> $ total\_pass := 0 $\\
\> $ total\_fail := 0$\\
\> \bc{ $error$ is a real in $[0,1]$.  It is the accumulated probability of all paths  }\ec\\
\> \bc{ which, because of the discretisation of the algorithm, we cannot determine exactly.}\ec\\
\> \bc{ This is where the revised version of the algorithm differs from the initial one.}\ec\\
\> \bc{ It is  initialised to zero. }\ec\\
\> \bc{ }\ec\\
\> $ error := 0$ \\
\> \bc{ $prob(s,t)$ is the probability of moving (from  anywhere) to location $s$ }\ec\\
\> \bc{  at time $t$. (i.e. in interval $(t-\delta,t]$.)}\ec\\
\> \bc{ For all combinations of locations and times, we initialise $prob$ }\ec\\
\> \bc{ to zero. }\ec\\
\> $\forall s \in S. \forall i \leq n$.\\
\>\>  $prob(s,\delta i) := 0$ \\
\> \bc{ $remain(s,t)$ is a boolean which is true if the probability of remaining }\ec\\
\> \bc{ in location $s$ during time interval $(t-\delta,t]$ is non-zero, false otherwise.}\ec\\
\> \bc{ They are all initialised to false.}\ec\\
\> $\forall s \in S. \forall i \leq n$.\\
\>\>  $remain(s,\delta i) := false$ \\
\> \bc{ $live(t)$ is the set of locations ``active'' at the end of }\ec\\
\> \bc{ interval $(t-\delta, t]$, which }\ec\\
\> \bc{ we need for calculating the information for the next time interval. }\ec\\
\> \bc{ For all time values, we initialise $live$ to the emptyset. }\ec\\
\>  $\forall i \leq n$.\\
\>\> $live(\delta i) := \emptyset$ \\
\> \bc{ We initialise all values in all matrices to zero.}\ec\\
\> \bc{ The are $n_s$ clocks in location $s$.}\ec\\
\> $\forall s \in S.$ \\
\>\> $\forall 0 \leq j \leq n.$ \\
\>\>\> $\forall 1 \leq i_1 \leq N_1$ \\
\>\>\>\>\> $\vdots$ \\
\>\>\> $\forall 1 \leq i_{n_s} \leq N_{n_s}. matrix(s,\delta j)[i_1\ldots i_{n_s}] := 0$\\
\\
\> \bc{ call procedure for calculating probabilities of clocks falling in the ranges }\ec\\
\> \bc{ $(0,\delta], (\delta,2\delta]$ etc.  This comes directly from the clock PDFs, }\ec\\
\> \bc{ and is only calculated once.  It is needed for determining the clock}\ec\\
\> \bc{probabilities.   }\ec\\
\> \bc{$C$ is the set of all clocks and  $F$ is the set of clock probability functions}\ec\\
\> \bc{ This procedure returns $pr$, which is needed in $new\_state\_matrix$ }\ec\\
\> \bc{ and $init\_matrix$. }\ec\\
\> $clock\_config\_probs(C,F,\delta,pr)$ \\
\> \bc{ }\ec\\
\\
\> \bc{ Section B}\ec\\
\> \bc{ Consider initial location of SA:  $s\_0$ }\ec\\
\> \bc{  If $s\_0 \models a\_1$ then formula is trivially true.  }\ec\\
\> if $s\_0 \models a_1$ then \\
\>\> $ total\_pass := 1 $\\
\> \bc{ If $s\_0 \models a\_0$ then formula is undecided and we   must   }\ec\\
\> \bc{  unfold SA further.  }\ec\\
\> elseif s\_0 $\models a_0$ then  \\
\>\> \bc{ Build the initial matrix, i.e. $matrix(s\_0,0)$.  }\ec\\
\>\> \bc{This will then contain the probabilities }\ec\\
\>\> \bc{of all the different clock settings for location $s\_0$ at time zero.   }\ec\\
\>\> $init\_matrix(matrix(s\_0,0))$  \\
\>\> \bc{ The only location ``live'' at time zero will be $s\_0$.  }\ec\\
\>\> $live(0) := \{s\_0\}$ \\
\> \bc{ If $s\_0$ does not model $a\_0$ or $a\_1$ then formula is trivially false.  }\ec\\
\> else \\
\>\> $total\_fail := 1$ \\
\> end if \\
\\
\> \bc{ Section C}\ec\\
\> \bc{ Each iteration of the following loop unfolds the automaton by }\ec\\
\> \bc{ one time step of $\delta$.  States which cause  the formula to  }\ec\\
\> \bc{ pass/fail are pruned from the tree, and their probabilities added to }\ec\\
\> \bc{ $total\_pass/total\_fail$, while the undecided states are recorded }\ec\\
\> \bc{ for the next iteration.  }\ec\\ 
\> \bc{  We continue while the values of $total\_pass$, $total\_fail$ and $error$ }\ec\\
\> \bc{ are not enough to  determine whether the formula is true or false }\ec\\
1: \>\> repeat \\
\>\> \bc{  Increment current\_time  }\ec\\
2:\>\> $ct := ct + \delta$ \\
\>\>\> \bc{  for all states $s$ that were live at the last clock tick }\ec\\
4:\>\>\> $\forall s \in live(ct - \delta)$ \\
\>\>\>\> \bc{  set current\_state to $s$.  }\ec\\
5:\>\>\>\> $cs := s$ \\
\>\>\>\> \bc{  The procedure $new\_time\_matrix$ returns }\ec\\
\>\>\>\> \bc{ $matrix(cs, ct)$: the matrix for the current state at the current time. }\ec\\
\>\>\>\> \bc{ It also   }\ec\\
\>\>\>\> \bc{ updates the function $prob$ with  the probability of remaining  }\ec\\
\>\>\>\> \bc{ in the current state at the current time and the  probabilities of   }\ec\\
\>\>\>\> \bc{ moving to  different states at the current time. }\ec\\
\>\>\>\>  \bc{ It also updates the value of $error$. }\ec\\
6:\>\>\>\> $new\_time\_matrix(matrix(cs, ct),new\_states(cs,ct), remain(cs,ct), prob, error)$ \\
\>\>\>\> \bc{ If the probability of remaining in current state at current time is zero  }\ec\\
7:\>\>\>\> if $remain(cs, ct) = false$ then \\
\>\>\>\>\> \bc{  current state is not  live at current time and  }\ec\\ 
\>\>\>\>\> \bc{  only the states which can be reached from current state at current time }\ec\\
\>\>\>\>\> \bc{  are added to those live at current time }\ec\\
8:\>\>\>\>\> $live(ct) :=  live(ct) \bigcup new\_states(cs,ct)$ \\
9:\>\>\>\> else \bc{ $remain(cs, ct) = true$ }\ec\\
\>\>\>\>\> \bc{  The current state, plus all states which may be reached from it at }\ec\\
\>\>\>\>\> \bc{ the current time, must be added to the live states. }\ec\\
10:\>\>\>\>\> $live(ct) := live(ct) \bigcup \{cs\} \bigcup  new\_states(cs,ct)$ \\
11:\>\>\>\>end if \\
11a:\>\>\> end forall \bc{ $\forall s \in live(ct - \delta)$ }\ec\\
\> \bc{ Now, we have $live(ct)$ and $prob(cs,ct)$ for all $cs$ in $live(ct)$ }\ec\\
\> \bc{ i.e. all the states we could be in at time $ct$, and the probability of  }\ec\\
\> \bc{ actually entering them in the previous time interval.  }\ec\\
\> \bc{ }\ec\\
\>\>\>\> \bc{ For every state which can be reached  at the current }\ec\\
\>\>\>\> \bc{ time, we must see if it causes the formula to pass or fail, in }\ec\\
\>\>\>\> \bc{ which cases we adjust the values for $total\_pass$ or }\ec\\
\>\>\>\> \bc{ $total\_fail$ and remove the state from the $live$ set.  If we cannot yet }\ec\\
\>\>\>\> \bc{ tell whether the formula is true or false, we must build the state/time matrix. }\ec\\
12:\>\>\>\> $\forall q \in live(ct)$  \\
\>\>\>\>\> \bc{  if  $q \models a_1$, then formula is true }\ec\\
13:\>\>\>\>\> if $q \models a_1$ then \\
\>\>\>\>\>\> \bc{ $total\_pass$ is incremented by the probability of entering $q$ }\ec\\
\>\>\>\>\>\> \bc{  from the current state at the current time }\ec\\
14:\>\>\>\>\>\> $total\_pass := total\_pass + prob(q,ct)$ \\
\>\>\>\>\>\> \bc{  State $q$ is removed from the live set }\ec\\
15:\>\>\>\>\>\> $live(ct) := live(ct) \setminus \{q\}$ \\
\>\>\>\>\> \bc{  Otherwise, if  $q \models a_0$ (and $q$ is not a terminating state)  }\ec\\
\>\>\>\>\> \bc{ then the formula may still be true, }\ec\\
\>\>\>\>\> \bc{ so we must build $matrix(q,ct)$ and keep state $q$ in the $live(ct)$ set. }\ec\\
16:\>\>\>\>\> elseif $q \models a_0 \land q \not\in terminating\_states$ then \\
\>\>\>\>\>\> \bc{ The procedure $new\_state\_matrix$ returns }\ec\\
\>\>\>\>\>\> \bc{ $matrix(q,ct)$: the matrix for state $q$ at current time, and requires }\ec\\
\>\>\>\>\>\> \bc{ $prob(q,ct)$: the probability of entering state $q$ from the current }\ec\\
\>\>\>\>\>\> \bc{ state at the current time. }\ec\\
17:\>\>\>\>\>\> $new\_state\_matrix(matrix(q, ct), prob(q,ct))$ \\
18:\>\>\>\>\> else  \bc{ If $q$ does not model $a\_0$ or it is a terminating state and also  }\ec\\
\>\>\>\>\>\>\> \bc{ it does not model $a\_1$ then the formula is false }\ec\\
\>\>\>\>\>\>\> \bc{ $total\_fail$ is incremented by the probability of entering $q$ }\ec\\
\>\>\>\>\>\>\> \bc{  from the current state at the current time }\ec\\
19:\>\>\>\>\>\> $total\_fail := total\_fail + prob(q,ct)$ \\
\>\>\>\>\>\> \bc{  State $q$ is removed from the live set }\ec\\
20:\>\>\>\>\>\> $live(ct) := live(ct) \setminus \{q\}$ \\
21:\>\>\>\>\> end if \\
22:\>\>\>\> end forall \bc{ for all states  in  $live(ct)$ }\ec\\
23:\>\> until $total\_pass > p$ \bc{ formula has passed }\ec\\
24:\>\>\>\> or \\
25:\>\>\>\> $total\_fail \geq 1-p$ \bc{ formula has failed }\ec\\
26:\>\>\>\> or \\
27:\>\>\>\> ($error \geq 1-p \land error \geq p$) \bc{ no possibility of a pass or a fail }\ec\\
28:\>\>\>\> or \\
29:\>\>\>\> ct = t \bc{ time's up.}\ec\\
30:\>\> if ($ct = t$) then \\
\>\>\> \bc{ All states undecided at the last iteration are now false, so }\ec\\
\>\>\> \bc{ $total\_fail$ is set to  $1 - total\_pass - error$ }\ec\\
31: \>\>\> $total\_fail :=  1 - total\_pass - error$\\
32:\>\> end if \\
\> \bc{}\ec\\
\> \bc{ Output result, based on the values of}\ec\\ 
\> \bc{ $total\_pass$, $total\_fail$ and $error$ }\ec\\
33:\>\> if $total\_pass  > p$ then \\
\>\> \bc{ SA models formula  }\ec\\
34:\>\> output pass  \\
35:\>\> elseif  \bc{ $total\_fail \geq 1-p$ }\ec\\
\>\> \bc{ SA does not model formula }\ec\\
36:\>\> output fail \\
37:\>\> else \bc{ errors are too large; cannot decide }\ec\\
38:\>\> output undecided \\
39:\>\> end if \\
\end{tabbing}

\begin{tabbing}
....\=....\=....\=....\=....\=....\=....\=....\=....\=....\=....\=....\=
\kill

\bc{ This procedure builds  the initial matrix. }\ec\\
\bc{ We assume there are $n$ clocks associated with this state, }\ec\\
\bc{ and $c^{s_0}_l$ is the $l$th clock. }\ec\\
\bc{ We abbreviate $\lceil upper\_bound(c^{s_0}_l)\rceil.\frac{1}{\delta}$ by $N_l$. }\ec\\

\\
$procedure \; init\_matrix(matrix(s_0,0))$ \\
begin procedure \\
\> $\forall 1 \leq i_1 \leq N_1$ \\
\>\>\> $\vdots$ \\
\> $\forall 1 \leq i_n \leq N_n . matrix(s_0,0)[i_1\ldots i_n] := {\displaystyle \prod_{l=1}^{n}} pr(c^{s_0}_l \in [i_l-\delta,i_l))$ \\
end procedure \\
\end{tabbing}

\begin{tabbing}
....\=....\=....\=....\=....\=....\=....\=....\=....\=....\=....\=....\=
\kill
$procedure \; new\_time\_matrix(matrix(cs,ct),new\_states(cs,ct),remain(cs,ct),prob,error)$ \\
\> \bc{ This procedure updates  a matrix by incrementing time, not by }\ec\\
\> \bc{ changing state.  We can do this by considering the values in the previous time }\ec\\
\> \bc{ matrix.   It also updates the function $prob$,}\ec\\
\> \bc{ and the variable $error$.}\ec\\
\> \bc{ There are $n$ clocks in state $cs$.}\ec\\
begin procedure \\
1:\> $\forall 1 \leq i_1 \leq N_1$ \\
\>\>\> $\vdots$ \\
2:\> $\forall 1 \leq i_{n} \leq N_{n} .$ \\
\>\>\>\>\>\>\> \bc{  If one of the matrix indices is at its maximum value, then the }\ec\\
\>\>\>\>\>\>\> \bc{ probability  value in this position must be zero.  This is }\ec\\
\>\>\>\>\>\>\> \bc{ because this procedure is always the first to update a state/time matrix. }\ec\\
\>\>\>\>\>\>\> \bc{  }\ec\\
\>\>\>\>\>\>\> \bc{  }\ec\\
3:\>\>\>\>\>\>\> if $\exists l \leq n \bullet i_l = N_l$ then \\
4:\>\>\>\>\>\>\>\> $matrix(cs,ct)[i_1,\ldots , i_{n}] := 0$ \\
\>\>\>\>\>\>\>\>  \bc{ otherwise the values in the matrix can be updated  simply from the }\ec\\
\>\>\>\>\>\>\>\>  \bc{ values in the previous time matrix.  }\ec\\
5:\>\>\>\>\>\>\> else \bc{ all clocks $c_i$ are $\geq 1$ and $<N_i$ }\ec\\
6:\>\>\>\>\>\>\>\> $matrix(cs,ct)[i_1,\ldots ,i_{n}] := $\\
7:\>\>\>\>\>\>\>\>\>\> $matrix(cs,ct)[i_1,\ldots ,i_{n}] + matrix(cs,ct-\delta)[i_1{\scriptstyle +1},\ldots ,i_{n} {\scriptstyle +1}]$ \\
\>\>\>\>\>\>\>\> \bc{ we record the fact that it is possible to remain in this state  }\ec\\
\>\>\>\>\>\>\>\> \bc{ at this time. }\ec\\
8:\>\>\>\>\>\>\>\> $remain(cs,ct):= true$ \\
9:\>\>\>\>\>\>\> end if \\
9a:end forall \\
\> \bc{ We now pick out the positions in the previous time matrix which, }\ec\\
\> \bc{ when moved forward one unit in time, result in a new state.  }\ec\\
10:\> $\;\forall 1 \leq i_1 \leq N_1$ \\
\>\>\> $\vdots$ \\
11:\> $\;\forall 1 \leq i_{n} \leq N_{n}$ \\
\> \bc{ If more than one of the previous time matrix indices is one,  we know that }\ec\\
\> \bc{  more than one of the clocks will have reached zero by $ct$, and so we  }\ec\\
\> \bc{  add the probability to error.  }\ec\\
11a:\>\>\>\>\>\>\> if $\#\{ c_l \mid c_l = 1\} > 1$ then \\
12:\>\>\>\>\>\>\>\> $error := error + matrix(cs,ct-\delta)[i_1, \ldots , i_n]$ \\
12a:\>\>\>\>\>\>\> else if  $\#\{ c_l \mid c_l = 1\} = 1$\\
\>\>\>\>\>\>\>\> \bc{   Given the stochastic Automaton $SA$, the state $cs$ and the clock $cc$ }\ec\\
\>\>\>\>\>\>\>\> \bc{  $s'$ is the  resulting state.  If the clock is associated with more than  }\ec\\
\>\>\>\>\>\>\>\> \bc{  one transition the function $pick$ (the adversary) chooses the }\ec\\
\>\>\>\>\>\>\>\> \bc{ resulting state.  Otherwise the state is the one determined by the  }\ec\\
\>\>\>\>\>\>\>\> \bc{ transition relation of the SA.  }\ec\\
13:\>\>\>\>\>\>\>\> $s' := pick(SA,cs,c_l)$ \\
13a:\>\>\>\>\>\>\>\> $new\_states(cs,ct) := new\_states(cs,ct) \bigcup \{s'\}$ \\
\>\>\>\>\>\>\>\> \bc{  the probability of entering $s'$  at time $ct$ }\ec\\
\>\>\>\>\>\>\>\> \bc{ is incremented by the matrix probability }\ec\\ 
14:\>\>\>\>\>\>\>\> $prob(s',ct) := prob(s',ct) + matrix(cs,ct-\delta)[i_1,\ldots , i_{n}]$ \\ 
22:\>\>\>\>\>\>
\> end if \bc{line 11}\ec\\
23:\>\>\>\>\>\> end forall \\
24:end procedure \\
\end{tabbing}

\begin{tabbing}
....\=....\=....\=....\=....\=....\=....\=....\=....\=....\=....\=....\=
\kill

\bc{ This procedure builds a new matrix, where the state is new rather than the time }\ec\\
\bc{ We assume there are $n$ clocks associated with this state, }\ec\\
\bc{ and $c^{s}_l$ is the $l$th clock. }\ec\\
\bc{ We abbreviate $\lceil upper\_bound(c^{s}_l)\rceil.\frac{1}{\delta}$ by $N_l$. }\ec\\
\bc{ The values in the matrix are calculated by  multiplying the clock }\ec\\
\bc{ probabilities by a factor of $p$, where $p$ is the probability of }\ec\\
\bc{ entering the state, and adding this value to the value already in }\ec\\
\bc{ the position.  }\ec\\

\\
$procedure \; new\_state\_matrix(matrix(cs,ct),p)$ \\
begin procedure \\
\> $\forall 1 \leq i_1 \leq N_1$ \\
\>\>\> $\vdots$ \\
\> $\forall 1 \leq i_{n} \leq N_{n} . matrix(cs,ct)[i_1,\ldots ,i_{n}] := $\\
\>\>\>\>\>\>\>\>\>\> $matrix(cs,ct)[i_1,\ldots ,i_{n}] + (p \times {\displaystyle \prod_{l=1}^{n}} pr(c^{s}_l \in [i_l-\delta,i_l))\;)$ \\
end procedure \\
\end{tabbing}

%

 \end{document}